\documentclass[a4paper,11pt]{article}
\pdfoutput=1
\usepackage{jheppub}
\usepackage{slashed}
\usepackage[T1]{fontenc}
\usepackage{adjustbox}
\usepackage{multirow}
\usepackage{hhline}
\usepackage{tabularx,booktabs,caption}
\usepackage{float}
\usepackage{placeins}
\usepackage{caption}
\usepackage{verbatim}
\usepackage{soul}
\usepackage[utf8]{inputenc}

\newcommand{\expt}[1]{\left\langle #1 \right\rangle}
\newcommand{\ie}{$i.e.$}
\newcommand{\eg}{$e.g.$}
\newcommand{\nn}{\nonumber{\nonumber}}
\newcommand{\abs}[1]{\left| #1 \right|}
\newcommand{\beq}{\begin{equation}}
\newcommand{\eeq}{\end{equation}}
\newcommand{\beqn}{\begin{eqnarray}}
\newcommand{\eeqn}{\end{eqnarray}}
\newcommand{\bea}{\begin{eqnarray}}
\newcommand{\eea}{\end{eqnarray}}

\title{\boldmath Freeze-in Dark Matter from Secret Neutrino Interactions}
\preprint{ACFI-T20-04, UCI-HEP-TR-2020-09}

\author[a]{Yong Du,}
\author[b,c]{Fei Huang,}
\author[b]{Hao-Lin Li}
\author[b,d,e]{and Jiang-Hao Yu} 

\affiliation[a]{Amherst Center for Fundamental Interactions, Physics Department,\\
University of Massachusetts Amherst, Amherst, MA 01003 USA}
\affiliation[b]{CAS Key Laboratory of Theoretical Physics, Institute of Theoretical Physics,\\
Chinese Academy of Sciences, Beijing 100190, China}
\affiliation[c]{Department of Physics and Astronomy, University of California, Irvine, CA 92697 USA}
\affiliation[d]{School of Physical Sciences, University of Chinese Academy of Sciences, Beijing 100049, P.\ R.\ China}
\affiliation[e]{School of Fundamental Physics and Mathematical Sciences, Hangzhou Institute for Advanced Study, University of Chinese Academy of Sciences, Hangzhou 310024, China}

\emailAdd{yongdu@umass.edu}
\emailAdd{huangf4@uci.edu}
\emailAdd{lihaolin@itp.ac.cn}
\emailAdd{jhyu@itp.ac.cn}

\abstract{We investigate a simplified freeze-in dark-matter model in which the dark matter only interacts with the standard-model neutrinos via a light scalar. The extremely small coupling for the freeze-in mechanism is naturally realized in several neutrino-portal scenarios with the secret neutrino interactions. We study possible evolution history of the hidden sector: the dark sector would undergo pure freeze-in production if the interactions between the dark-sector particles are negligible, while thermal equilibrium within the dark sector could occur if the reannihilation of the dark matter and the scalar mediator is rapid enough. We investigate the relic abundance in the freeze-in and dark freeze-out regimes, calculate evolution of the dark temperature, and study its phenomenological aspects on BBN and CMB constraints, the indirect-detection signature, as well as the potential to solve the small scale structure problem.}

\begin{document} 
\captionsetup[figure]{labelfont={bf},labelformat={default},labelsep=period,name={FIG.}}
\maketitle

\section{Introduction}\label{sec:intro}
Nowadays, the existence of dark matter has been firmly supported by numerous astrophysical and cosmological observations.
However, despite indisputable evidence for its existence, the nature of dark matter remains a mystery.
Indeed, all the observational evidence, such as the observations on the galactic rotation curve \cite{Rubin:1970zza}, the bullet cluster \cite{Clowe:2006eq} and the cosmic microwave background (CMB) \cite{Aghanim:2018eyx}, depends on the gravitational interaction of dark matter only.
Due to the lack of knowledge on its non-gravitational interaction,
the properties of dark matter, such as its mass, spin, its couplings to the Standard Model (SM), and whether it is composite or fundamental, remain unknown, nor do we know anything about its production mechanism.
Understanding the nature of dark matter would not be so urgent if it plays an insignificant role in the universe.
However, dark matter occupies roughly a quarter of the total energy density of the universe today, which is about five times the energy density of the content from the Standard Model (SM),
it is the dominating species that governs the expansion of the universe in a large portion of the cosmic history, and it seeds the formation of structure which gives rise to the galaxies and the clusters of galaxies that we observe today.
Therefore, resolving the mystery of dark matter is one of the biggest challenges facing particle physics, astrophysics and cosmology today.

Among the numerous attempts to tackle the problem of dark matter, the weakly interacting massive particle (WIMP) is perhaps the most studied dark-matter candidate.
In the standard WIMP paradigm, the mass and the interaction strength between the WIMPs and the SM particles are often assumed to be at the weak scale. 
The WIMPs are assumed to be in thermal equilibrium with the radiation bath in the radiation dominated (RD) epoch, but freeze out of the thermal equilibrium when the interaction rate falls below the expansion rate.
It is often referred to as the WIMP miracle that the combination of weak scale mass and coupling strength together with the thermal freeze-out mechanism successfully gives rise to the correct dark-matter relic abundance, $\Omega_{\rm DM}~\simeq~0.26$.
However, aside from its minimality and strong motivation from many models of the physics beyond the Standard Model (BSM), such as theories of supersymmetry, decades of null result in both direct and indirect detections, as well as collider searches, have made the standard WIMP scenarios more and more constrained \cite{Roszkowski:2017nbc}.
Therefore, the current situation has invoked a lot of interest in other dark-matter scenarios.

{
Indeed, by modifying some of the defining features of the thermal WIMP paradigm, other viable dark-matter scenarios could also be constructed.
For example, dark matter need not be produced via the thermal freeze-out mechanism.
In fact, the interaction between dark matter and the SM particles could be very feeble such that the thermal equilibrium is never established between dark matter and the SM thermal bath.
An equally motivated production mechanism suitable for this type of scenarios is the freeze-in mechanism \cite{Hall:2009bx,Elahi:2014fsa,Bernal:2017kxu,Krnjaic:2017tio,Berger:2018xyd} in which dark matter is produced gradually by the injection from the visible thermal bath.
Typically, if dark matter is coupled to the thermal bath via renormalizable operators, its production is IR dominated by the temperature near its mass and thus insensitive to initial conditions.
On the other hand, for non-renormalizable interactions \cite{Elahi:2014fsa}, the production is sensitive to UV physics. 
Another approach is to modify the single-particle picture of most WIMP models.
For instance, models from the non-minimal dark-sector framework often depend on the collective behavior of an entire ensemble of dark-sector particles \cite{Dienes:2011ja,Dienes:2011sa,Dienes:2016vei}.
One could also modify the standard $\Lambda$ cold dark matter ($\Lambda$CDM) cosmology and consider dark-matter production in non-standard cosmologies, such as producing thermal dark matter during a matter dominated universe before the Big-Bang Nucleosynthesis (BBN), see \cite{Kane:2015jia, Acharya:2009zt} for reviews and \cite{Berlin:2016vnh,Berlin:2016gtr,Heurtier:2019eou} for recent progress.
By considering different possibilities, the relevant constraints and the potential signatures could be drastically different.}

In this paper, we are motivated by the possibility that the interaction between the dark matter and the SM thermal bath is very feeble such that the dark-matter particles could only be populated gradually via the annihilation of the SM particles.
To be specific, we consider a \emph{hidden sector} consisting of a dark scalar mediator $\phi$ and a fermionic dark matter $\chi$, and the hidden sector is only connected to the SM via a neutrino portal.
Typically, there are two kinds of neutrino-portal dark-matter models --- the $t$-channel models with the $\nu-\chi-\phi$ interaction \cite{Falkowski:2009yz,Macias:2015cna,Ko:2015nma,Hamze:2014wca,Yu:2016lof, Gonzalez-Macias:2016vxy,Escudero:2016tzx,Escudero:2016ksa,Batell:2017rol,Batell:2017cmf,Chianese:2018dsz,Bian:2018mkl,Blennow:2019fhy,Chianese:2019epo}, and the $s$-channel models with the $\nu-\nu-\phi$ interaction \cite{Cherry:2014xra,Berlin:2018ztp}, in the freeze-out scenarios. 
While the $t$-channel models have been investigated in details in recent years,
in this paper, we consider the $s$-channel models in which the dark matter $\chi$ only interacts with the SM neutrinos through an s-channel exchange of $\phi$ in the freeze-in regime.

{The neutrino portal could naturally give rise to a tiny coupling between $\phi$ and the SM neutrinos in the low-energy effective theory, which is suitable for the hidden-sector freeze-in production of dark matter.
Although the dark matter can never thermalize with the SM thermal bath in freeze-in scenarios, we shall see that the dynamics within the dark sector might be able to significantly affect the dark-matter production.
In particular, if the interaction rate between $\chi$ and $\phi$ is negligible, the dark-matter production depends only on the energy injection from the SM thermal bath.
However, if such interaction is sufficiently rapid, the reannihilation of $\chi$ and $\phi$ could establish a period of thermal equilibrium within the dark sector \cite{Cheung:2010gj,Cheung:2010gk,Chu:2011be,Bernal:2015ova,Bernal:2017kxu,Krnjaic:2017tio,Berger:2018xyd}.
In the latter case, the dark-sector particles would eventually break away from the thermal equilibrium either by directly freezing out of the dark thermal equilibrium as their interaction rate drops below the expansion rate, 
or by entering a phase called the \emph{Quasi-Static equilibrium} (QSE) \cite{Cheung:2010gj} in which the annihilation rate of $\chi$ is much larger than that of $\phi$ and is balanced with the injection from the SM sector before finally freezing out of it.
Ultimately, the relic abundance of the dark matter is sensitive to the differences in the evolution history.}

A sizable interaction strength within the dark sector would allow rapid annihilations of $\chi$ inside dark-matter halos at the present epoch.
Such late-time annihilations of $\chi$ followed by a rapid decay of $\phi$ could produce a substantial amount of neutrinos and thus would be constrained by the current neutrino observations.
In addition, a sufficiently large coupling between $\chi$ and $\phi$ could be exploited to solve problems on small scale structure at late time,
making our model a potentially viable candidate for the self-interacting dark matter (SIDM, see \eg~Ref.~\cite{Tulin:2017ara} for review and the references in it).

This paper is organized as follows. 
In Sec.\,\ref{sec:model}, we set up the neutrino-portal dark-matter model and discuss several examples for UV completion. 
In Sec.\,\ref{sec:evo}, we study the evolution of the dark sector in different regimes and discuss the consequence of having different evolution history.
In Sec.\,\ref{sec:phe}, we discuss the phenomenological aspects of our model, including the constraints on the decay of the dark scalar $\phi$, the potential for our model to solve the small scale structure problems, and the indirect-detection constraints. 
Finally, we conclude in Sec.\,\ref{sec:con}.

\section{The Neutrino Portal Model}\label{sec:model}
In the $s$-channel neutrino model, we consider a dark sector which consists of a Dirac fermionic $\chi$ and a light scalar mediator $\phi$.
While the dark-matter candidate $\chi$ is directly coupled to $\phi$ only, the mediator $\phi$ is in addition coupled to the SM neutrinos.
To be specific, the effective Lagrangian for a dark sector described above can be in general written as
{
\beqn
{\cal L}&=&{\cal L}_{\rm kin}+{\cal L}_{\rm int}\,\,\,,\\
{\cal L}_{\rm kin}&=&\frac{1}{2}(\partial_\mu \phi\partial^\mu \phi-m_\phi^2\phi^2)+\bar{\chi}(i\slashed{\partial} -m_{\chi})\chi\,\,\,,\\
{\cal L}_{\rm int}&=&-\phi \bar{\chi}(g_\chi+i\bar{g}_\chi\gamma^5) \chi-\phi \bar{\nu}_i(g_\nu^{ij}+i\bar{g}_\nu^{ij}\gamma^5)\nu_j \label{eq:lag_int}\,\,\,,\label{eq:lag_general}
\eeqn}
{
where $g_{\chi}$ and $\bar{g}_{\chi}$ are the couplings between the mediator $\phi$ and the dark matter $\chi$, while $g_{\nu}^{ij}$ and $\bar{g}_{\nu}^{ij}$ are the effective couplings of $\phi$ to the SM neutrinos in the flavor basis with the superscript $i$ and $j$ being the flavor index.}
In our model, we do not assume any non-vanishing vacuum expectation value (VEV) for the $\phi$ field.
Consequently, 
1) $\phi$ does not mix with the SM Higgs boson and could thus evade all the collider constraints; 
2) the only way for the dark sector to interact with the SM sector is through the light mediator $\phi$. 
However, it is possible that the Higgs VEV might contribute to the mass of $\phi$, which we implicitly include in the $m_\phi$ term.

To produce dark matter through freeze-in in the early universe, we generically require the couplings {$g_\nu^{ij}$ and $\bar{g}_\nu^{ij}$} to be very tiny in order to avoid establishing thermal equilibrium between the dark and the SM sectors.
Such small couplings could be naturally generated through appropriate UV models. 
In what follows, we shall discuss the UV completion for the general Lagrangian above in two limiting scenarios
-- the one with a pure scalar $\phi$ and only pure-scalar interactions; and the one in which $\phi$ is a pseudoscalar, and only pseudoscalar interactions exist.
We shall refer to these two scenarios as {\it Scenario I} and {\it Scenario II}. 
The interacting Lagrangian in Eq.~(\ref{eq:lag_int}) can be rewritten by setting $\bar{g}_{\chi(\nu)}~=~0$ or $g_{\chi(\nu)}~=~0$ for the two scenarios, respectively:
{
\bea
{\cal L}_{\rm int}&=&\left\{
\begin{array}{ll}-g_\chi\phi \bar{\chi}\chi - g_\nu^{ij} \phi \bar{\nu}_i \nu_j & ({\rm Scenario~ I}),\\
-i\bar{g}_\chi\phi \bar{\chi}\gamma^5 \chi - i\bar{g}_\nu^{ij} \phi \bar{\nu}_i \gamma^5 \nu_j & ({\rm Scenario~ II}).
\end{array} \label{eq:lag_eff}
\right.
\eea}

For Scenario I, 
we are motivated by the type-I seesaw and introduce the right-handed neutrinos $N$ in addition to the scalar $\phi$.
We assume the masses $M_{N}~\gg~M_{\rm EW}$ with $M_{\rm EW}$ being the electroweak scale. 
At high energy scale, the Lagrangian can be written as
\bea
\mathcal{L}_{\rm int}~\supset~g_{H} \bar{L} \tilde{H} N+g_{\phi}\overline{N^c}N\phi + h.c.\,\,\,,
\eea
where $\tilde{H}\equiv i\sigma_2H^*$. 
Around the electroweak scale, one can integrate out the heavy fields $N_i$ and obtain the dimension-6 operators:\footnote{We shall suppress the neutrino flavor index from now on, if not otherwise mentioned.}
\bea
	\mathcal{L}_{\rm int}~\supset~\frac{1}{M_{N}^2} {\phi (\bar{L}\tilde{H}) (\tilde{H}^T L^c)} \,\,\,.
\eea
After electroweak spontaneous symmetry breaking, tree level interactions between the scalar $\phi$ and the SM neutrinos can be generated as follows:
\bea
	\frac{1}{M_{N}^2} {\phi (\bar{L} \tilde{H}) (\tilde{H}^\dagger L)} \to g_\nu \phi \bar{\nu} \nu,\quad  {\textrm{~with~}} g_\nu \sim \frac{v^2}{M_N^2},
\eea
where $v$ is the SM Higgs VEV. 
Note that since we assume $M_{N}~\gg~M_{\rm EW}$, $g_\nu$ is naturally suppressed.

For Scenario II, the $\phi-\nu-\nu$ interaction for a pseudo-scalar $\phi$ could be generated in the so-called minimal Majoron model\,\cite{Chikashige:1980qk,Chikashige:1980ui,Schechter:1981cv,Gelmini:1980re}.
In this type of model, we still introduce three right-handed neutrinos $N_i$ but in addition a SM-singlet complex scalar field $\Phi$ above the electroweak scale. 
At high energy scale, the interacting Lagrangian can be written as
\bea
\mathcal{L}_{\rm int}~\supset~g_{\Phi} \overline{N^c}N\Phi - g_H{\overline{L}} \tilde{H} N + h.c.\,\,\,.\label{eq:majoron1}
\eea
We then assume that $\Phi$ develops a VEV and spontaneously breaks a global $U(1)$ symmetry at a scale higher than $M_{\rm EW}$.
Therefore, we can write $\Phi=\varphi+v_\Phi+i\phi$, where $v_{\Phi}$ is the VEV, the pseudoscalar $\phi$ is the Nambu-Goldstone boson, and we assume $v_{\Phi}~\gg~v$.
We further assume that the $U(1)$ symmetry is not exact, but explicitly broken by a soft-breaking mass term in the potential of $\Phi$.
As a result, the pseudoscalar $\phi$ acquires a small mass and becomes a pseudo Nambu-Goldstone boson.
After the symmetry breaking, first term in Eq.~(\ref{eq:majoron1}) generates the Majorana mass term and the interacting Lagrangian becomes
\bea
\mathcal{L}_{\rm int}~\supset~-\frac{M_N}{2}\bar{N}N-ig_\Phi \phi\bar{N}\gamma_5 N - g_H{\overline{L}} \tilde{H} N \,\,\,,
\eea
with $M_N~=~g_\Phi v_\Phi$, where we have transform $N$ to the four component notation for a Majorana field such that $N^c=N$.
We can then integrate out the heavy filed $N$, and the effective Lagrangian can be expressed as
\bea
{\cal L}_{\rm int}~\supset~ i\frac{{g}_\Phi g_H^2}{M_N^2} \phi ({\overline{L}} \tilde{H}) \gamma^5 (\tilde{H}^T L^c ) + h.c.\,\,\,.
\eea 
Finally, after the electroweak spontaneous symmetry breaking,
\bea
{\cal L}_{\rm int}~\supset~ - i\bar{g}_\nu\phi \bar{\nu} \gamma^5 \nu\,,\quad {\text{with~~}} \bar{g}_\nu~\sim~\frac{v^2}{M_N^2}
~\sim~\frac{v^2}{g_\Phi^2v_\Phi^2}\,\,\,.
\eea 
Again, $\bar{g}_\nu$ is naturally suppressed. 
We point out that it has also been studied when the pNGB $\phi$ becomes a dark-matter candidate in Refs.\,\cite{Gelmini:1984pe,Lattanzi:2013uza,Berezinsky:1993fm,Bazzocchi:2008fh,Gu:2010ys,Abe:2020dut}.

\begin{figure}\centering
\includegraphics[width=1.\textwidth]{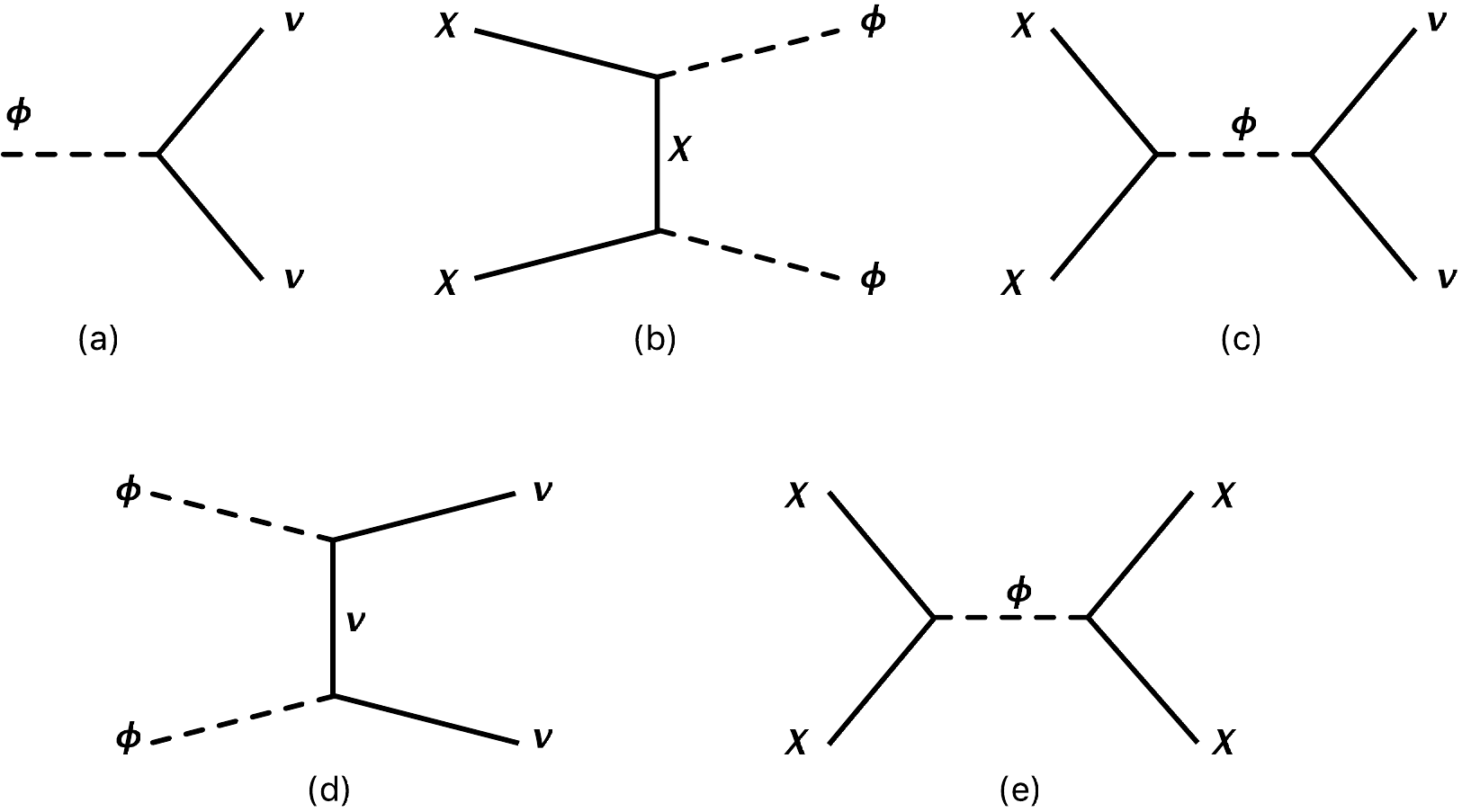}
\caption{Relevant Feynman diagrams in our study.}\label{feyndia}
\end{figure}

Having demonstrated how Scenario I and Scenario II could be generated from UV theories, in what follows, we shall simply use the effective Lagrangian in Eq.~(\ref{eq:lag_eff}) when discussing production of dark matter.
Note that, in the above discussion, we choose not to specify the coupling strength between $\chi$ and $\phi$.
As a consequence, while a highly suppressed $g_{\nu}$ or $\bar{g}_{\nu}$ could ensure a sufficiently feeble interaction between the SM sector and the dark sector during the dark-matter production, by tuning the couplings $g_{\chi}$ or $\bar{g}_{\chi}$, we are able to study interesting dark-sector dynamics and phenomenologies in several different regimes.
To conclude this section, we list all the Feynman diagrams relevant for our scenarios in FIG.\,\ref{feyndia}.

\FloatBarrier
\section{The Freeze-In Production of Dark Matter}\label{sec:evo}
As we have discussed in the previous sections, we are interested in the regime where the interaction between the dark-matter particles and the SM neutrinos is too feeble to ever establish thermal equilibrium between them, and the dark sector is mainly populated by a gradual injection of energy from the SM thermal bath, \ie~via the freeze-in mechanism. 
{Since the couplings in Eq.~(\ref{eq:lag_general}) are all renormalizable, our scenario falls in the category of IR freeze-in.
Similar to the freeze-out scenario, it is well know that the final yield of dark-matter particles is insensitive to the initial condition in IR freeze-in scenarios.}
Therefore, without loss of generality, we shall also assume that the initial energy density of the dark-sector particles is negligible. To be concrete, the evolution of dark-matter number density is determined by the following Boltzmann equation
\beqn
\frac{dn_\chi}{dt}
&=&-3Hn_\chi-\left(n_{\chi}^2-n_{\chi}^{\rm eq}(T)^2\right) \expt{\sigma_{\chi\chi\to\nu\nu} v}_{T} - n_{\chi}^2 \expt{\sigma_{\chi\chi\to\phi\phi} v}_{T_{\chi}}
+n_{\phi}^2 \expt{\sigma_{\phi\phi\to\chi\chi} v}_{T_{\phi}}\,\,\,,\nn\\
\label{eq:Boltzmann_nchi}
\eeqn
where $T$ refers to the temperature of the SM thermal bath, and $T_{\chi,\phi}$ are the temperatures that the dark-sector particles $\chi$ and $\phi$ might have during their evolution which are not necessarily identical.
In the equation above, we have implicitly assumed that the phase-space distribution of $\chi$ and $\phi$ takes the form $f_{\chi,\phi}~\sim~e^{-(E-\mu_{\chi,\phi})/T_{\chi,\phi}}$ with their own temperature and chemical potential.
Besides, the condition of detailed balance, $f^{\rm eq}_\nu(p_{\nu_1})f^{\rm eq}_\nu(p_{\nu_2})=f^{\rm eq}_\chi(p_{\chi_1})f^{\rm eq}_\chi(p_{\chi_2})$, is applied to convert $n_{\nu}^{\rm eq}(T)^2\expt{\sigma_{\nu\nu\to\chi\chi}}_T$ --- the term that describes the production of dark matter from neutrino annihilation,
 into $n_{\chi}^{\rm eq}(T)^2\expt{\sigma_{\chi\chi\to\nu\nu}}_T$ which involves only the number density of $\chi$ at the temperature of the SM thermal bath.
In general, if the scattering rates are too slow, $\phi$ or $\chi$ might not be able to reach kinetic equilibrium --- 
they might not have a thermal distribution with a well defined temperature.
Nevertheless, we shall keep using the thermally averaged cross-section assuming that it is a good approximation when the typical kinetic energy of the dark-sector particles is $T_{\chi,\phi}$. 

As it is already discussed in Ref~\cite{Cheung:2010gj,Cheung:2010gk,Chu:2011be,Krnjaic:2017tio,Berger:2018xyd}, depending on $g_\chi$ and/or $\bar{g}_\chi$ and the dark-sector number densities, the production of dark-sector particles can be categorized into two different regimes: 
\begin{itemize}
\item The \emph{pure freeze-in regime}, where the couplings between $\chi$ and $\phi$ are so small that interactions between them are negligible. As a result, $\chi$ and $\phi$ can never reach thermal equilibrium.\\ 
\item The \emph{reannihilation regime}, where the couplings $g_\chi$ and/or $\bar{g}_\chi$ are large enough that the annihilations process $\chi\chi\leftrightarrow\phi\phi$ starts to play a significant role in the evolution of $n_\chi$ and $n_\phi$. This could also allow $\chi$ and $\phi$ to reach thermal equilibrium during a certain period of time in the evolution of the universe with a common temperature denoted as $T_D$ which is smaller than the temperature of the SM thermal bath $T$.
\end{itemize}

\begin{figure}\centering
\includegraphics[width=0.9\textwidth]{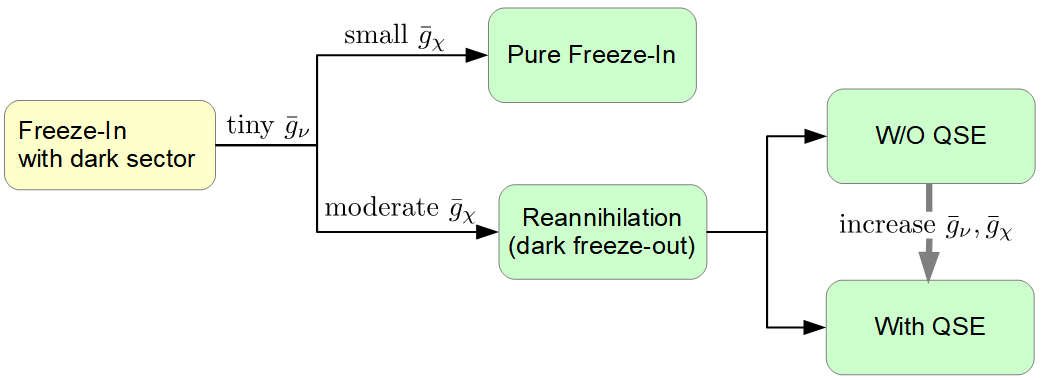}
\caption{ The classification of production mechanism for our freeze-in dark matter model. $\bar{g}_\chi$ and $\bar{g}_\nu$ are the pseudo-scalar couplings of the mediator $\phi$ to the dark matter particle $\chi$ and the SM neutrino respectively. 
The quasi-static equilibrium (QSE)\cite{Cheung:2010gj} refers to the situation in which the $\chi\chi\to\phi\phi$ process greatly exceeds its inverse process due to the injection of $\chi$ from the thermal bath and the injection of $\chi$ is balanced by its annihilation into $\phi$.
}\label{fg:fchart}
\end{figure}

In what follows, we describe these two regimes in detail while focusing on Scenario II in which $\phi$ is a pseudoscalar case, as the discussion can be easily carried over to Scenario I in which $\phi$ is a pure scalar. We also associate the FIG.~\ref{fg:fchart} for a summary for the different regime with the corresponding scale of the relevant couplings. 

\subsection{Pure Freeze-in Regime}
As mentioned above, in the pure freeze-in regime, not only the dark matter is decoupled from the SM thermal bath, the thermal equilibrium between dark-sector constituents $\chi$ and $\phi$ is also never reached.
Therefore, during the production of dark-matter particles $\chi$,
one can make the following approximations: 
1) the number density $n_{\chi}$ is so small that the annihilation process $\chi\chi\rightarrow \nu\nu$ can be safely neglected;
2) the particle exchange between dark-sector species $\phi\phi \leftrightarrow \chi\chi$ is also negligible.
With these approximations, the Boltzmann equation Eq.~(\ref{eq:Boltzmann_nchi}) can be approximated as
\beqn
\frac{dn_\chi}{dt}
~\approx~-3Hn_\chi+ n_{\chi}^{\rm eq}(T)^2\langle \sigma_{\chi\chi\to\nu\nu} v \rangle_{T}\,\,\,,\label{eq:Boltzmann_pure_fi}
\eeqn
where $T$ represents the temperature of the SM thermal bath, and we have used the condition $n_{\chi}^{\rm eq}(T)~\gg~n_{\chi}(T)$. 

To connect with the observed dark-matter relic density today, it is often convenient to use entropy conservation $\tilde{s}a^3~=~\rm const.$ where $\tilde{s}$ is the entropy density and $a$ is the scale factor, 
and define the comoving number density or the \emph{yield} of dark matter \mbox{$Y_{\chi}~\equiv~n_{\chi}/\tilde{s}$} which eventually becomes a constant after the freeze-in process concludes.
Therefore, to reproduce the expected present-day dark-matter relic density, the final yield $Y_{\chi}(T_f)$ is required to be 
\beq
Y_{\chi}(T_f)=Y_{\chi}(T_{\rm now})\approx \frac{\rho_{\rm crit}\Omega_{\chi}}{m_{\chi}s(t_{\rm now})}\approx4.14\times 10^{-10}\times\left(\frac{\rm GeV}{m_{\chi}}\right)\,\,\,.\label{eq:yield_purefi}
\eeq
The above expression can thus be used to constrain our model parameters.
In what follows, we shall estimate the yield from the pure freeze-in process.
We first note that
\beqn
{\sigma_{\chi\chi\rightarrow\nu\nu}v_{\text{M\o l}}} 
&=& \frac{3\bar{g}_{\chi}^2\bar{g}_{\nu}^2 s^2}{64\pi [(s-m_\phi^2)^2+m_\phi^2\Gamma_\phi^2]E_{\chi_1}E_{\chi_2}} \,\,\,,\label{eq:xxphiphiv}
\eeqn
where $\Gamma_\phi$ is the decay width of $\phi$, $s$ is the Mandelstam variable, and $E_{\chi_{1}},~E_{\chi_{2}}$ represent the energy of the dark-matter particles in the annihilation process.
In the high-temperature regime ($T~\gg~m_{\chi}$), most of the annihilations take place with $s~\gg~4m_{\chi}^2$, whereas in the low-temperature regime ($T~\ll~m_{\chi}$), typically $s~\sim~4m_{\chi}^2$.
Taking these approximations, the thermally averaged cross-section can be expressed as follows assuming $m_\chi~\gg~m_\phi$:
\[
\expt{\sigma_{\chi\chi\rightarrow\nu\nu}v_{\text{M\o l}}}_T
~\approx~\left\{
\begin{array}{ll}
\displaystyle\frac{3\bar{g}_{\chi}^2\bar{g}_{\nu}^2 }{128\pi \langle E_\chi\rangle^2}\simeq 7.5\times 10^{-4}\times\frac{\bar{g}_{\chi}^2\bar{g}_{\nu}^2}{T^2}\,\,\,,&\quad T\gg m_{\chi}\\
\displaystyle\frac{3\bar{g}_{\chi}^2\bar{g}_{\nu}^2 }{64\pi \langle E_\chi\rangle^2}\simeq 0.015\times  \frac{\bar{g}_{\chi}^2\bar{g}_{\nu}^2}{m_{\chi}^2}\,\,\,, &\quad T\ll m_{\chi}
\end{array}
\right.\,,
\label{eq:chichi2phiphiv}
\]
where we have applied $\expt{E_{\chi}}~=~{\rho_{\chi}}/{n_{\chi}}~\simeq~3.15\,T$ when $T~\gg~m_{\chi}$ and $\expt{E_{\chi}}~\simeq~m_{\chi}$ when $T\ll m_{\chi}$. 
Assuming that $f_{\chi}~=~\exp(-E_\chi/T)$, we further obtain the expressions for the annihilation rate: 
\beqn
n_{\chi}^{\rm eq} \expt{\sigma_{\chi\chi\rightarrow\nu\nu} v}_T~\simeq~\left\{ 
\begin{array}{ll} 1.4 \times 10^{-4}~\bar{g}_{\chi}^2\bar{g}_{\nu}^2~T\,\,\,, & \quad T\gg m_{\chi} \\
1.9\times10^{-3}~\bar{g}_{\chi}^2\bar{g}_{\nu}^2~T\times e^{-m_{\chi}/T}\sqrt{\displaystyle\frac{T}{m_{\chi}}}\,\,\,, & \quad T\ll m_{\chi} 
\end{array} 
\right.\,.\label{eq:ann_rate}
\eeqn
In the pure freeze-in regime, since the $\nu\nu\to\chi\chi$ process is the only production channel for the dark-matter particles , the condition that the dark matter never reaches thermal equilibrium with the SM generally requires that
\beq
n_{\chi}^{\rm eq} \expt{\sigma_{\chi\chi\rightarrow\nu\nu} v}_T~<~H\,,\label{eq:condition_neq}
\eeq
where $H$ is the Hubble expansion rate. 
In the radiation dominated epoch, we have
\beq
H~\simeq~\sqrt{\frac{\rho_R}{3M_P^2}}~=~\sqrt{\frac{\pi^2}{90}}~g_{\star}^{1/2}(T)\frac{T^2}{M_P}\,\,\,,\label{eq:H_RD} 
\eeq
where $g_\star$ is the effective numbers of relativistic degrees of freedom for the energy density, and $M_P$ is the reduced Planck mass. 
Since the number density in Eq.~(\ref{eq:ann_rate}) is always exponentially suppressed in the non-relativistic regime, we simply require Eq.~(\ref{eq:condition_neq}) to be satisfied in the relativistic regime, which implies
\beqn
\bar{g}_{\chi}\bar{g}_{\nu} ~\lesssim~ \mathcal{O}(10^{-8})~\left(\frac{m_{\chi}}{\rm GeV}\right)^{1/2}\,\,\,,\label{eq:fi_condition}
\eeqn
if $g_{\star}^{1/2}(T)~\sim~\mathcal{O}(10)$.

To proceed further with the estimate, we rewrite the Eq.~(\ref{eq:Boltzmann_pure_fi}) as
\beqn
\frac{dY_{\chi}}{dT}~\approx~-\frac{s\expt{\sigma v}}{HT}Y_{\chi}^{\rm eq}(T)^2\,\,\,,
\eeqn
which we can directly integrate from a large initial temperature $T_i~\gg~m_{\chi}$ to the present-day temperature $T_{\rm now}$:
\beqn
\int_{Y_{\chi}^i}^{Y_{\chi}^{\rm now}} dY_{\chi}
~\approx~-\int_{T_{\rm now}}^{T_i} dT~\frac{s\expt{\sigma v}}{HT}Y_{\chi}^{\rm eq}(T)^2 \,\,\label{eq:int_yield}
\eeqn
Using our previous discussion, the integrand
\begin{equation}
\frac{s\expt{\sigma v}}{HT}Y_{\chi}^{\rm eq}(T)^2~\propto~\left\{
\begin{aligned}
& 1/T^2\,\,, & \quad T~\gg~m_\chi \\
& e^{-2m_\chi/T}/T^3\,\,, &  \quad T~\ll~m_\chi 
\end{aligned}
 \right.\,\,.
\end{equation}
Obviously, the Boltzmann suppression factor which appears in the non-relativistic regime suggests that the contribution to the entire integral from this part, \ie~\mbox{$m_{\chi}>~T~\geq~T_{\rm now}$}, is always subleading.
At the same time, the $1/T^2$ behavior in the relativistic regime also indicates that the production of dark matter is very small when $T~\gg~m_{\chi}$.
In fact, most of the contribution to the integral in Eq.~(\ref{eq:int_yield}) actually comes from the part where $T~\sim~m_\chi$.
Therefore, we find as expected that the dark-matter production in the pure freeze-in regime is largely insensitive to the initial conditions.

Based on our discussions, we can then make following approximations and roughly estimate the final yield:
\beq
{Y_{\chi}^{\rm now}}~\approx~\int^{m_{\chi}}_{T_i} dT~\frac{s\expt{\sigma v}}{HT}Y_{\chi}^{\rm eq}(T)^2~\approx~2.1\times 10^{14}\times \frac{\bar{g}_{\chi}^2\bar{g}_{\nu}^2}{g_{\star}^{1/2}(m_{\chi})g_{\star,s}(m_{\chi})}\left(\frac{\rm GeV}{m_{\chi}}\right)\,\,\,,\label{eq:yield_approx}
\eeq
in which $g_{\star,s}$ is the effective numbers of relativistic degrees of freedom for entropy density.
For $m_{\chi}~\sim~\mathcal{O}(\rm GeV)$, the denominator $g_{\star}^{1/2}g_{\star,s}~\sim~10^3$.
Comparing with Eq.~(\ref{eq:yield_purefi}),
one generally need $\bar{g}_{\chi}\bar{g}_{\nu}~\sim~\mathcal{O}(10^{-11})$ to obtain the correct dark-matter relic abundance.

\begin{figure}\centering
\includegraphics[width=0.8\textwidth]{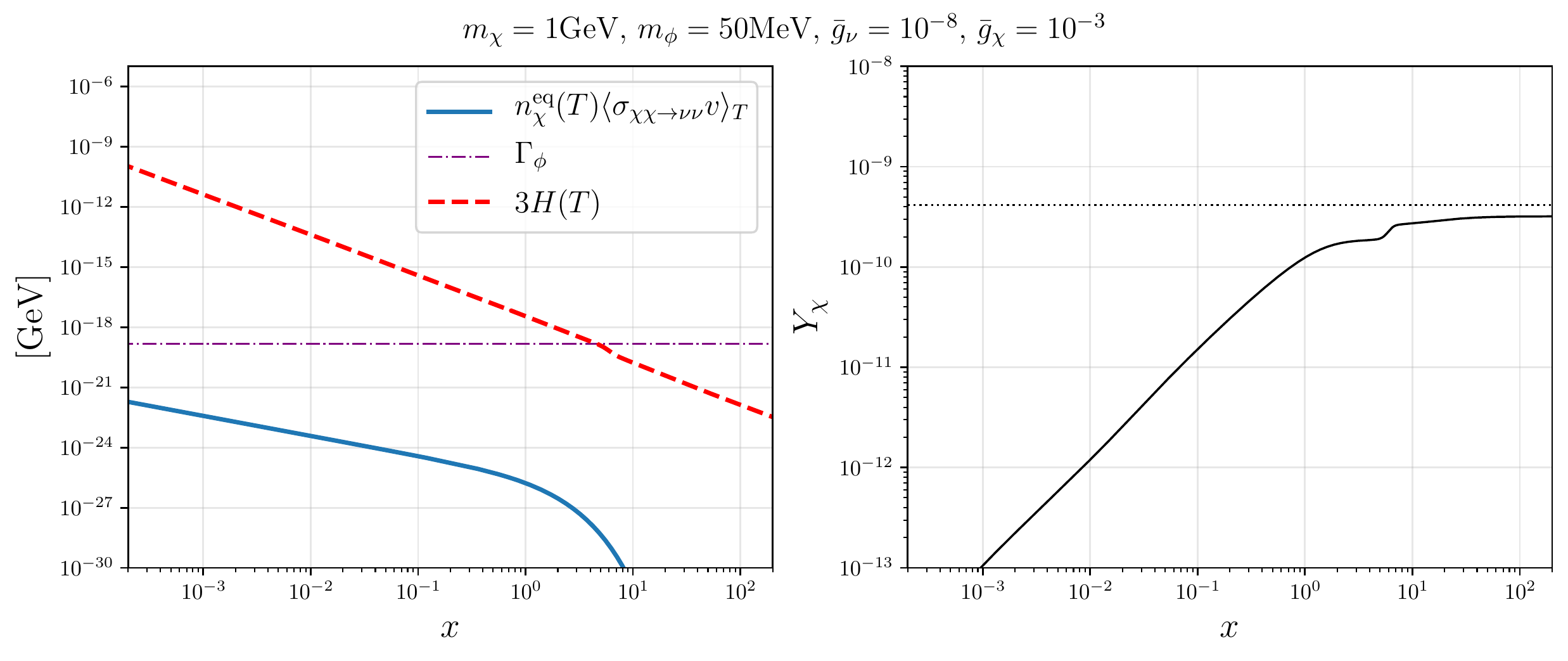}\\
\includegraphics[width=0.8\textwidth]{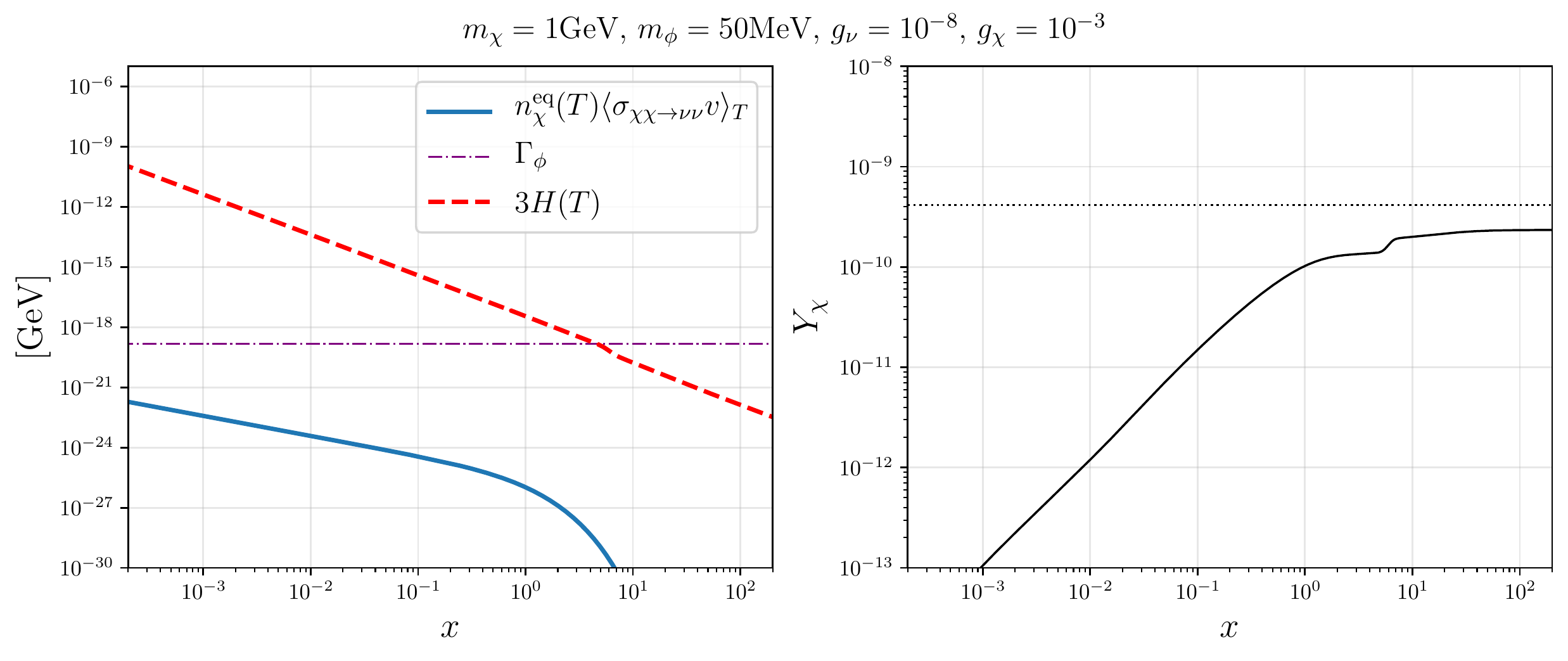}
\caption{Examples in the pure freeze-in regime.
The upper panels correspond to \emph{Scenario II} where $\phi$ is a pseudoscalar, and the lower panels correspond to \emph{Scenario I} where $\phi$ is a pure scalar.
In the left panels, the blue curve shows the evolution of the dark-matter production rate $n_{\chi}^{\rm eq}(T)\expt{\sigma_{\chi\chi\to\nu\nu}v}_T$, whereas the dash-dotted purple line is the decay rate of $\phi$.
The dashed red curve is the evolution of the Hubble expansion rate $H$.
In the right panels, we show the evolution of $Y_{\chi}$ as a function of $x~\equiv~m_{\chi}/T$. 
The dotted horizontal line indicates the observed relic density of the dark matter today. 
}\label{fg:purefi}
\end{figure}

Examples of the pure freeze-in scenario for both the pseudoscalar (upper panels) and scalar (lower panels) cases are shown in FIG.~\ref{fg:purefi}, in which we have chosen to plot against $x~\equiv~m_\chi/T$.
In the left panels, we show a comparison between the expansion rate $H$ and the dark-matter production rate $n_{\chi}^{\rm eq}(T)\expt{\sigma_{\chi\chi\to\nu\nu}v}_T$.
As expected, the dark-matter production rate is always smaller than $H$.
However, despite the fact that it is always decreasing as the SM temperature drops, it is closest to $H$ at $T~\sim~m_\chi$, which supports the approximation we use in Eq.~(\ref{eq:yield_approx}).
After the temperature drops below $m_{\chi}$, a quick down turn appears as the neutrinos no longer have enough energy to create the dark-matter particles.
As a result, we see in the right panel that the yield grows monotonically from a vanishingly small initial value, and then asymptotes to its maximum near $T~\sim~m_\chi$.
Note that the sudden jump for both $H$ and $Y_{\chi}$ around $x~\sim~5-10$ is not a numerical artifact but merely due to the sudden change in $g_{\star}$ and $g_{\star,s}$ when the temperature drops to $T~\sim~\mathcal{O}(100)~\rm MeV$, \ie,~when the QCD phase transition occurs.

Comparing the pseudoscalar with the pure scalar cases, it is obvious that for couplings of the same size, the final yield in the latter is slightly smaller.
This is simply because the factor $s^2$ in the numerator of the thermally averaged cross-section in Eq.~(\ref{eq:xxphiphiv}) becomes $s(s-4m_{\chi}^2)$ in Scenario I (see Eq.~(\ref{eq:xsection_xxnunu}) for more details), which results in a suppression in the production rate as $T$ approaches $m_{\chi}$.
\FloatBarrier

\subsection{Reannihilation Regime}

In the pure freeze-in regime, the interactions between the dark-sector particles are so weak such that both $\chi$ and $\phi$ barely annihilate. As a consequence, the $\chi\chi\leftrightarrow\phi\phi$ process has little impact on the evolution of dark-matter number density.
However, it is also possible that while the entire dark sector is decoupled from the SM sector, the annihilations of $\chi$ and $\phi$ proceed with a non-negligible rate.
Indeed, since the interaction strength between $\chi$ and $\phi$ is governed by the coupling $\bar{g}_{\chi}$ in Scenario II (or $g_{\chi}$ in Scenario I),
by increasing $\bar{g}_{\chi}$ while adjusting $\bar{g}_{\nu}$ accordingly in order to prevent thermalizing with the SM sector, as required in Eq.~(\ref{eq:fi_condition}),
the dark sector could establish its own thermal equilibrium for a period of time in the history of the universe as it is populated by the SM thermal bath.
This thermal equilibrium is characterized by a common temperature $T_D$ for both $\chi$ and $\phi$.
In this situation, Eq.~(\ref{eq:Boltzmann_nchi}) can be rewritten as
\beqn
\frac{dn_\chi}{dt}
&\approx&-3Hn_\chi+n_{\chi}^{\rm eq}(T)^2 \expt{\sigma_{\chi\chi\to\nu\nu} v}_{T}
-\left(n_{\chi}^2-n_{\chi}^{\rm eq}(T_D)^2\right)\expt{\sigma_{\chi\chi\to\phi\phi} v}_{T_D}\,\,\,,\label{eq:Boltzmann_dseq}
\eeqn
in which the the thermally averaged cross-section for $\chi\chi\leftrightarrow\phi\phi$ is evaluated at $T_D$.

The thermal equilibrium in the dark sector requires the interaction rate between the dark-sector particles 
\mbox{$n_{\chi}^{\rm eq}(T_D)\expt{\sigma_{\chi\chi\to\phi\phi}}_{T_D}>H(T)$}.
For a sizable cross-section $\sigma_{\chi\chi\to\phi\phi}$ , this condition could be immediately realized if the initial particle number density $n_{\chi}$ or $n_\phi$ is large enough.
By contrast, if the initial number density is small, the dark sector could start without thermal equilibrium but slowly build it up as the SM sector keeps populating the dark-sector particles.
Once the thermal equilibrium is established within the dark sector,
one would expect the number density $n_{\chi}$ to track $n_\chi^{\rm eq}(T_D)$, since the last term in the Boltzmann equation above always tends to bring back any deviation from the thermal equilibrium.
This is indeed true if the second term on the right-hand side of Eq.~(\ref{eq:Boltzmann_dseq}) is not large enough such that the $\chi\chi\to\phi\phi$ process can always be balanced by the $\phi\phi\to\chi\chi$ process.
However, if this SM source term is sufficiently large, $n_{\chi}$ might be able to grow significant larger than $n_\chi^{\rm eq}(T_D)$.
Therefore, the annihilation rate of $\chi$ could overwhelm that of $\phi$, and this could occur even when the latter is still larger than the Hubble rate.
In this situation, Eq.~(\ref{eq:Boltzmann_dseq}) can be approximated as 
\beqn
\frac{dn_\chi}{dt}
~\approx~-3Hn_\chi+n_{\chi}^{\rm eq}(T)^2 \expt{\sigma_{\chi\chi\to\nu\nu} v}_{T}
 -n_{\chi}^2\expt{\sigma_{\chi\chi\to\phi\phi} v}_{T_D}\,\,\,.\label{eq:Boltzmann_qse}
\eeqn
The equation above simply means that while the SM thermal bath is always sourcing the dark matter $\chi$, the annihilation of $\chi$ always tends to deplete this particle injection.
As a result, the QSE phase often occurs in which the last two terms in Eq.~(\ref{eq:Boltzmann_dseq}) are balanced against each other \cite{Cheung:2010gj}.
To be precise, the QSE phase is defined by
\beqn
Y_{\chi}(T)~\approx~Y_{\chi}^{\rm QSE}(T)~\equiv~Y_{\chi}^{\rm eq}(T)\sqrt{\frac{\expt{\sigma_{\chi\chi\to\nu\nu} v}_{T}}{\expt{\sigma_{\chi\chi\to\phi\phi} v}_{T_D}}}\,\,\,.
\eeqn

The dark thermal equilibrium or the QSE persists until $n_{\chi}(T)\expt{\sigma_{\chi\chi\to\phi\phi}}_{T_D}~\lesssim~H(T)$, and then the dark matter experiences a \emph{dark freeze-out}.
However, unlike the standard thermal freeze-out scenario, the comoving number density of the dark matter might not be fixed at its value around the dark freeze-out, because the production of $\chi$ from the SM thermal bath could still make a significant contribution to the final yield even after the dark freeze-out. 
In any case, differences caused by having different initial number densities for $\chi$ or $\phi$ would eventually be washed out by the thermalization within the dark sector --- the freeze-in mechanism is also independent of initial conditions in the reannihilation regime.

As mentioned above, Eq.~(\ref{eq:Boltzmann_dseq}) depends on the dark-sector temperature $T_D$, which, unlike the SM temperature, is not known \emph{a priori}.
In order to find $T_D$, we numerically solve the Boltzmann equation for the total energy density in the dark sector:
\beqn
\frac{d\rho_D}{dt}&=&-3H(P_D+\rho_D)+2n_\nu^{\rm eq}(T)^2\mathcal{P}_{\nu\nu\to\chi\chi}(T)+2n_\nu^{\rm eq}(T)^2\mathcal{P}_{\nu\nu\to\phi\phi}(T)
\nn\\
&~&+n_\nu^{\rm eq}(T)^2\mathcal{P}_{\nu\nu\to\phi}(T)
-n_\phi\mathcal{P}_{\phi\to\nu\nu}(T_D)\,\,\,,
\label{eq:rhodBoltzmann}
\eeqn
where the energy density $\rho_D\equiv\rho_\chi+\rho_\phi$, the pressure $P_D\equiv P_\chi+P_\phi$,
and the thermally averaged energy transfer rate $\mathcal{P}_{a\dots\to i\dots}$ gives the energy transferred to or from the dark sector through the process $a+\dots\to i+\dots$. 
We shall show its explicit form in Appendix~\ref{sec:energy_transfer}.
Note that the factor of 2 in the second and third term on the right-hand side takes into account the fact that two dark-sector particles are produced via each of the corresponding processes.
Again, we have made the assumption that the annihilation rate of the dark-sector particles into the SM neutrinos is negligible.
Such assumption is valid as long as $T_D\ll T$, which implies the number density of either $\chi$ or $\phi$ is much smaller than that of the SM neutrinos, and is also consistent with the second law of thermodynamics --- the dark sector can never be hotter than the SM thermal bath which populates it in the first place.
We shall see later that the decay and inverse decay could potentially thermalize $\phi$ with the SM neutrinos.
However, this has little impact for the yield of $\chi$, if any.

While in thermal equilibrium, the dark-sector energy density is only a function of 
the dark-sector temperature $T_D$ once the masses of $\chi$ and $\phi$ are fixed.
Therefore, at every step of the numerical simulation, $T_D$ can be obtained by solving
\beq
\rho_D(T_D)~=~\xi_\chi\int\frac{d^3p}{(2\pi)^3}\frac{E_\chi }{e^{E_\chi/T_D}+1}+\xi_\phi\int\frac{d^3p}{(2\pi)^3}\frac{E_\phi }{e^{E_\phi/T_D}-1}\,\,\,,\label{eq:rhod}
\eeq
in which $\xi_{\chi}=2$ and $\xi_{\phi}=1$ are the number of degrees of freedom for $\chi$ and $\phi$.
After obtaining the time dependence of the dark sector temperature $T_D(t)$, one can then insert it back to Eq.~(\ref{eq:Boltzmann_dseq}) and solve for $n_{\chi}$.

\begin{figure}\centering
\includegraphics[width=0.9\textwidth]{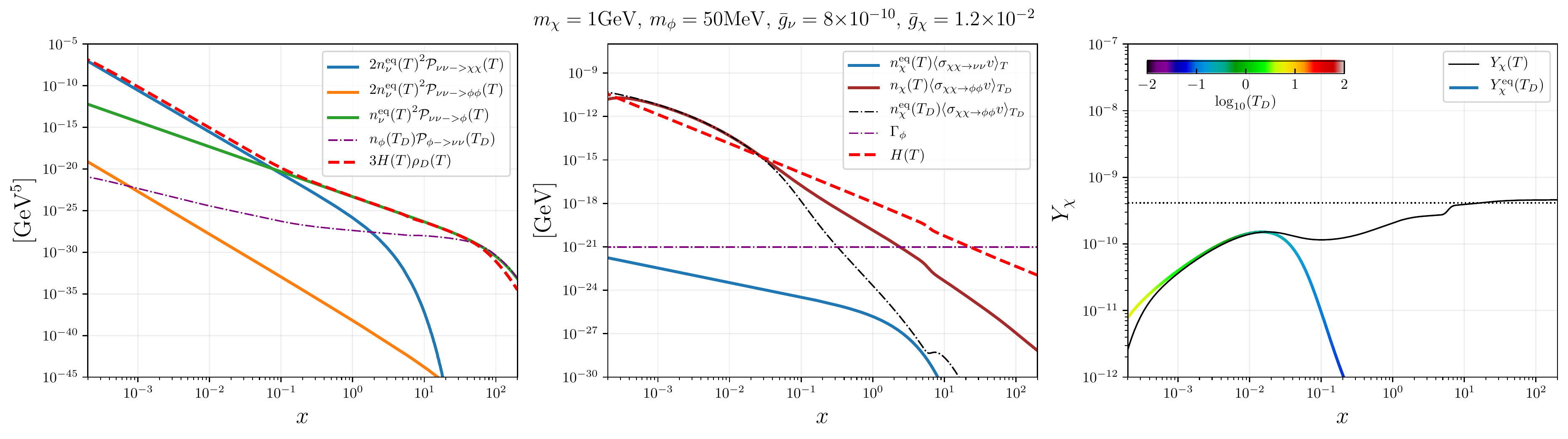}
\includegraphics[width=0.9\textwidth]{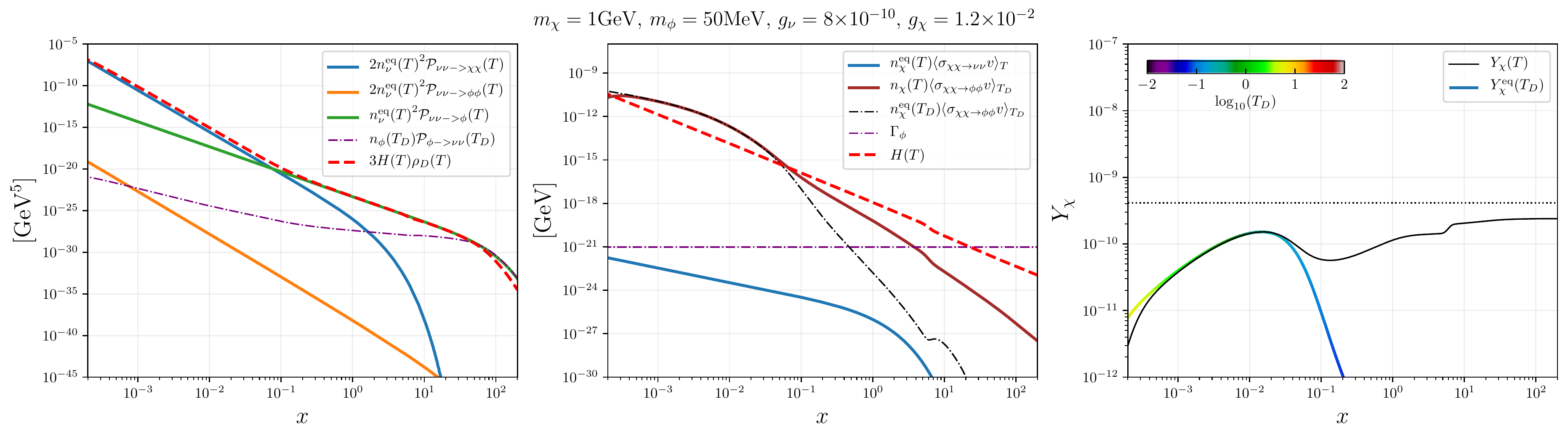}
\includegraphics[width=0.9\textwidth]{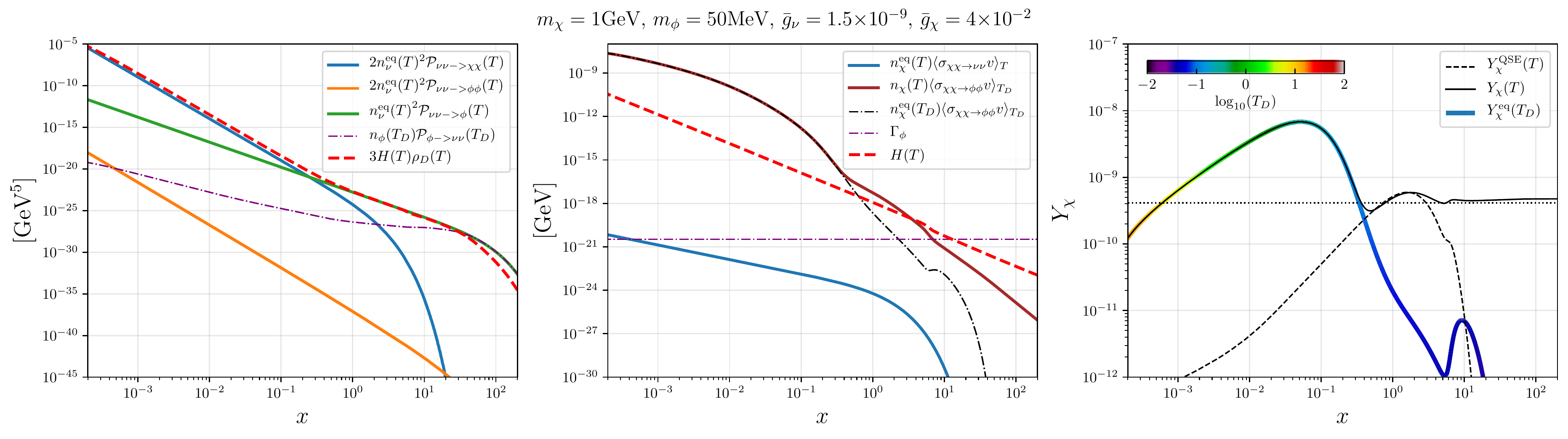}
\includegraphics[width=0.9\textwidth]{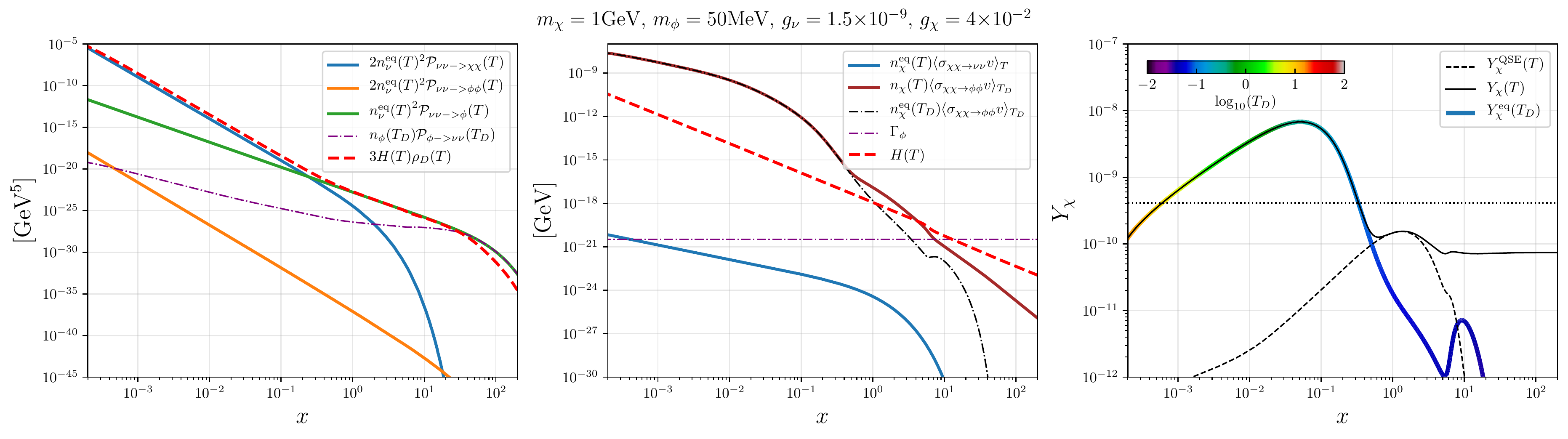}
\caption{Benchmark results for the reannihilation regime, with the
upper (lower) two panels showing the cases without (with) the QSE phase.
The left columns shows the rate of the energy transfer from the SM sector as well as the rates of energy-density dilution due to the expansion of the universe.
In the middle column, we compare the expansion rate with the interaction rates relevant for the dark-matter number density.
In the right column, we show the evolution of $Y_{\chi}$ and $Y_{\chi}^{\rm eq}(T_D)$ --- the yield of $\chi$ if it is in thermal equilibrium with $\phi$, which we have colored to indicate the corresponding $T_D$.
The evolution of $Y_{\chi}^{\rm QSE}(T)$) is also plotted in the lower two panels for the cases with the QSE phase.
}\label{fg:reannihilation}
\end{figure}

In FIG.~\ref{fg:reannihilation}, we present our benchmark results in the reannihilation regime including the evolution of the energy-transfer rates, the particle exchange rates, as well as the dark-matter yield.
The upper two panels show the cases in which the dark sector undergoes a period of dark thermal equilibrium and then freezes out.
By contrast, the cases with the additional QSE phase before the dark freeze-out are shown in the lower panels. 
Comparing with the pure freeze-in cases in FIG.~\ref{fg:purefi}, we have enhanced the coupling of $\phi$ to the dark matter, and at the same time lowered its coupling to the neutrinos in order to avoid thermalization with the SM thermal bath.
The couplings are tuned such that the cases in Scenario II could have a final yield of $\chi$ similar to that of the dark matter today.
At the same time, for cases with (or without) the QSE phase, we choose to use couplings of the same size just to demonstrate that the cases in Scenario I tend to have a smaller final yield due to the same reason that we have mentioned in the pure freeze-in cases.

In the left panels, we see for all the cases, most of the energy injected into the dark sector is from the $\nu\nu\to\chi\chi$ channel at early times (solid blue curve).
However, the contribution from this channel decreases drastically after the SM temperature drops below the mass of the dark-matter particles, as we can tell from the quick down turn of the solid blue curve for $x~\gtrsim~1$, and then the inverse decay $\nu\nu\to\phi$ (solid green curve) starts to dominate.
On the other hand, the energy transfer rate of the annihilation process $\nu\nu\to\phi\phi$ (solid orange curve), is always negligibly small. 
These observations are consistent with the expectations from our scenario in which $\bar{g}_{\nu}$ (or $g_{\nu}$) is very tiny, since the amplitude of the $\nu\nu\to\phi\phi$ process is proportional to $\bar{g}_{\nu}^2$ (or $g_{\nu}^2$), whereas the amplitudes of the $\nu\nu\to\chi\chi$ and $\nu\nu\to\phi$ processes are proportional to $\bar{g}_{\nu}\bar{g}_{\phi}$ and $\bar{g}_{\nu}$ (or $g_{\nu}g_{\phi}$ and $g_{\nu}$), respectively.

Comparing the cases with and without the QSE phase, we notice that
because both the coupling of $\phi$ to the dark matter and to the neutrinos are larger in the cases with the QSE phases, more energy is injected from the SM in these cases.
As a result, in the middle and the right panels, while the upper two cases only reach thermal equilibrium around $x~\sim~10^{-3}$, the lower two cases have already established thermal equilibrium long before that, and have acquired much more dark-matter particles when the dark sector is in thermal equilibrium. 
As a consequence, the evolution after reaching thermal equilibrium is also drastically different.

In the first two cases, $Y_{\chi}$ tracks $Y_{\chi}^{\rm eq}(T_D)$ until $n_{\chi}^{\rm eq}(T_D)\expt{\sigma_{\chi\chi\to\phi\phi}}_{T_D}~\lesssim~H$, which occurs around $x~\sim~0.02-0.05$.
At this time, the temperature in the dark sector $T_D$ is already smaller than $m_{\chi}$ as one can tell from the color variation in $Y_{\chi}^{\rm eq}(T_D)$.
As we have mentioned above, after the dark freeze-out, the yield $Y_{\chi}$ is not fixed yet.
Similar to the pure freeze-in scenario, the solid blue curve in the middle panel shows that the production of $\chi$ from the SM neutrinos is still significant until the SM temperature is much smaller than $m_{\chi}$.
Therefore, with the injection from the SM thermal bath continues to contribute, the yield only approaches its asymptote after $x~\gtrsim~1$.

In the bottom two cases, $Y_{\chi}$ tracks $Y_{\chi}^{\rm eq}(T_D)$ at the beginning, and can even grow larger than its final asymptotic value because of a larger $\bar{g}_{\nu}$ (or $g_\nu$).
These ``excess'' dark-matter particles are rapidly depleted by annihilating into $\phi$.
As a result, as $T_D$ approaches and goes below $m_{\chi}$, $Y_\chi$ grows increasingly slower and eventually starts to decrease (see the green part of $Y_{\chi}^{\rm eq}(T_D)$).
Note that, due to a larger coupling between $\chi$ and $\phi$, $Y_\chi$ can still track $Y_{\chi}^{\rm eq}(T_D)$ even when $\chi$ is already non-relativistic.
As $T_D$ keeps dropping, the inverse process $\phi\phi\to\chi\chi$ becomes less and less important due to kinematics.
However, since $\bar{g}_{\nu}$ (or $g_\nu$) is larger in these cases, the dark-matter annihilation rate $n_{\chi}\expt{\sigma_{\chi\chi\to\phi\phi}v}_{T_D}$ can still be larger than $H$.
This is clearly shown in the middle panels as the solid brown curve peels off from the dash-doted black curve.
Consequently, the dark matter breaks away from the thermal equilibrium and enters the QSE phase in which $Y_{\chi}$ tracks $Y_{\chi}^{\rm QSE}(T)$.
Finally, when $n_{\chi}(T)\expt{\sigma_{\chi\chi\to\phi\phi}v}_{T_D}~\lesssim~H(T)$, the QSE can no longer be maintained.
Then, $\chi$ freezes out, and the yield approaches its final value.

\subsubsection{\texorpdfstring{$T_D$}~~Evolution}
\begin{figure}[t]\centering
\includegraphics[width=0.6\textwidth]{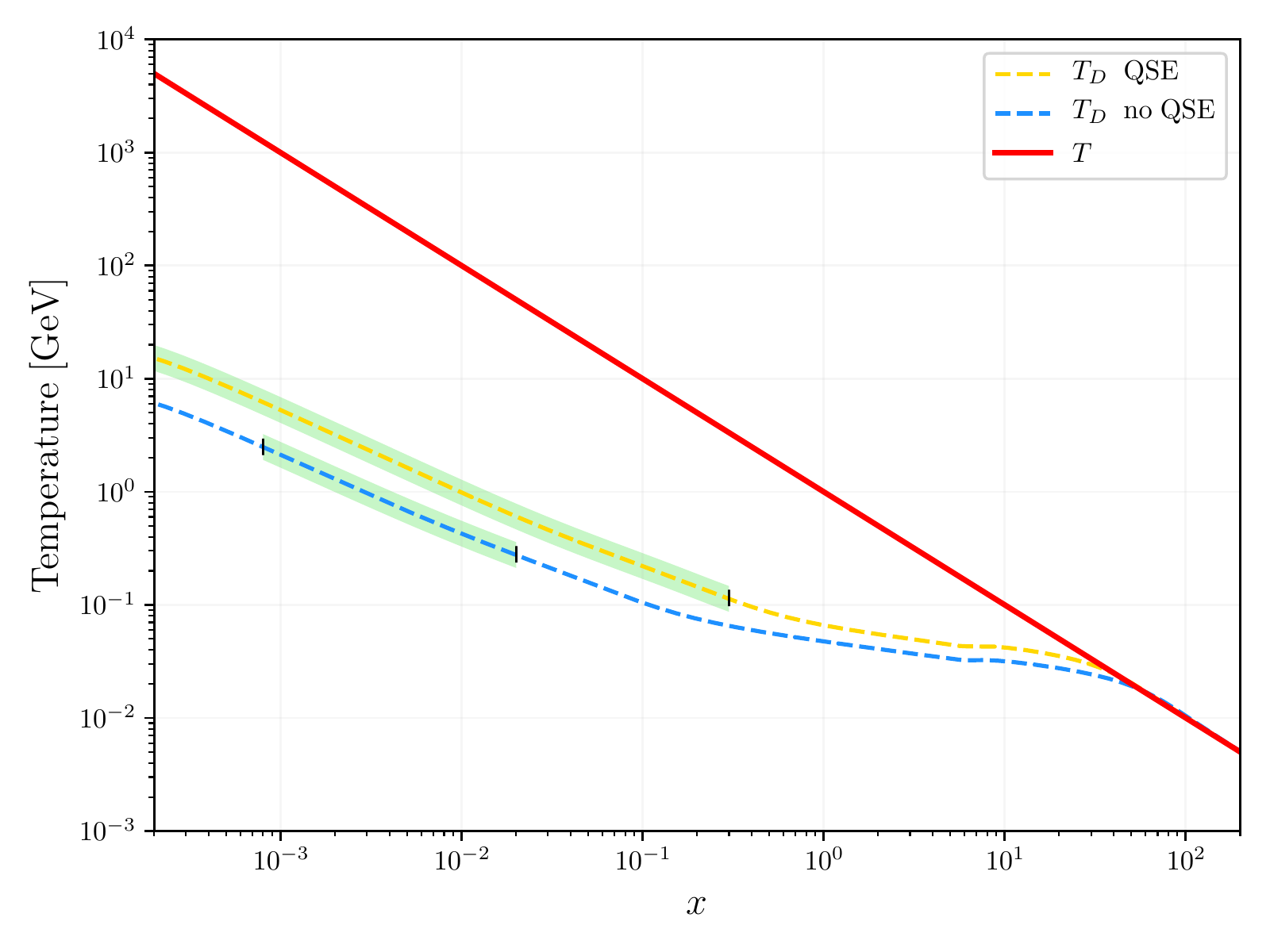}
\caption{The evolution of $T_D$ if the dark sector is in thermal equilibrium.
The yellow curve is obtained from the cases with the QSE phase, whereas the blue curve is from cases without it, and there is no visible distinction no matter $\phi$ is a scalar or pseudoscalar.
Since the temperature in the dark sector is only physical when $\chi$ and $\phi$ are in thermal equilibrium, we roughly shaded the region in which the dark thermal equilibrium is established.
As a reference, we use the red curve to show the temperature of the SM thermal bath.
}\label{fg:TD}
\end{figure}
We comment here on the evolution of dark-sector temperature $T_D$ in the the reannihilation regime.
The curves for $T_D$ which are obtained by solving Eq.~(\ref{eq:rhod}) are shown explicitly in FIG.~\ref{fg:TD}.
Note that, we only include two curves for $T_D$ because there is no visible distinction between the scalar and pseudoscalar cases.
It is worth emphasizing that, though the solution to Eq.~(\ref{eq:rhod}) is unique and exists in the entire regime we are considering, $T_D$ is only physically meaningful when the dark-sector particles $\chi$ and $\phi$ are in thermal equilibrium.
We therefore shade the region in which the dark thermal equilibrium is reached.
Within this region, we can see that slopes of both $T_D$ curves are less steep than the that of $T$.
This is because while the energy density of the SM thermal bath is only subject to the expansion of the universe and some minor impact from the change of $g_{\star}$, the dark-sector energy density is in addition always sourced by the SM.
Therefore, its temperature $T_D$ decreases slower than $T$.
After $\chi$ and $\phi$ decouple, the momentum distributions of $\chi$ and $\phi$ evolve independently.
As we can see from the left panels of FIG.~\ref{fg:reannihilation}, the inverse decay of neutrinos into $\phi$ starts to surpass the annihilation into a pair of $\chi$ around $x~\sim~0.1$, which is shortly before or after $\phi$ and $\chi$ decouple.
After that, most of the energy injection from the SM thermal bath goes into $\phi$.
However, the energy that goes into $\phi$ and cannot be rapidly converted into $\chi$ since $\chi$ is already non-relativistic when it decouples.
The energy density in the dark sector is thus dominated by $\phi$ after decoupling.
Therefore, assuming that the kinetic equilibrium is still maintained for $\phi$, and that chemical potential is negligible, the $T_D$ curves to the right of the shaded region largely reflect the evolution of the energy density of $\phi$, \ie~$T_D~\approx~T_{\phi}$ since $\rho_D~\approx~\rho_{\phi}^{\rm eq}(T_D)$.
However, $T_D$ always tends to be an overestimate for $T_{\phi}$ in this region as $\rho_{\chi}~>~\rho_{\chi}^{\rm eq}(T_D)$ after $\chi$ decouple.
Eventually, when $H(T)~\sim~\Gamma_{\phi}$, $\phi$ decays rapidly into neutrinos.
For the choice of $m_{\phi}$ and $\bar{g}_{\nu}$ (or $g_{\nu}$) in the examples, the decay occurs around $x~\sim~10-20$, when $\phi$ starts to transition from relativistic to non-relativistic.
Therefore, though both $T_D$ curves eventually merge with $T$, which indicates that $\phi$ could thermalizes with the SM sector through decay and inverse decay, its energy density is already suppressed at that time --- it is only the neutrinos living on the high momentum tail of the Fermi-Dirac distribution that are able to create the $\phi$ particles, which, in turn, quickly decay and deposit its energy back to the SM thermal bath.

\FloatBarrier

\section{Model Phenomenologies}\label{sec:phe}
In this section, we discuss general considerations on the phenomenological aspects of our model.
We notice that, at early times, the existence of the dark scalar $\phi$ might contribute to a non-negligible fraction of energy density in the universe, and its decay would inject energy into the SM thermal bath.
Therefore, our model could be constrained by measurements on BBN and CMB.
At late times, the interaction within the dark sector could allow a large annihilation rate inside dark-matter halos which would produce a lot of neutrinos.
Thus our model also receives constraints from the current observations on astrophysical neutrinos.
Moreover, the self-interaction between dark-matter particles could also suppress the formation of small structures.
As a result, our model could potentially be exploited to solve the small scale structure problems. 
In what follows, we discuss these aspects in turn based on the general effective Lagrangian in Eq.~(\ref{eq:lag_general}) instead of referring to any specific UV completion.

\subsection{Dark Scalar Decay}
The existence of $\phi$ and its decay might lead to measurable effects by changing the expansion rate of the universe and injecting SM neutrinos to the visible sector.

On one hand, $\phi$ could decay out of equilibrium, 
\ie~$\Gamma_\phi/H(m_\phi)~<~1$ (or equivalently $T_{\rm dec}~<~m_{\phi}$), where $T_{\rm dec}$ is the SM temperature when $H~=~\Gamma_\phi$.
If this occurs before BBN, it simply reheats the universe by producing entropy into the visible sector since neutrinos are still tightly coupled to the SM plasma.
Therefore, other than slightly changing the temperature evolution of the SM thermal bath and diluting the relic density of $\chi$, the decay of $\phi$ would not leave any measurable effect on BBN, even if $\phi$ has a significant amount of energy density before decaying.
However, if such out-of-equilibrium decay occurs during BBN but before the neutrino decoupling, it could potentially disturb the formation of light elements by changing the way the SM temperature evolves.
In addition, if $\phi$ decays after the neutrino decoupling, it could contribute to the effective number of neutrino species $N_{\rm eff}$ which can be constrained by CMB measurements as $\Delta N_{\rm eff}~\equiv~N_{\rm eff}-3.045$ \cite{Tanabashi:2018oca}.
Therefore, to avoid upsetting BBN and CMB, we basically need $\phi$ and its decay products to have a negligible contribution to the total energy density of the universe when $T~\lesssim~10 \rm MeV$.

On the other hand, if $\Gamma_\phi/H(m_\phi)~>~1$ (or $T_{\rm dec}~>~m_{\phi}$), the decay and inverse decay would bring $\phi$ into thermal equilibrium with the SM neutrinos.
If this occurs before neutrino decoupling, $\phi$ then becomes an additional component of the visible plasma and could contribute as much as one relativistic degree of freedom to the total energy density when it is relativistic.
This could lead to significant increase of the expansion rate during BBN, which would affect the prediction for the abundances of light elements.
If $\phi$ decays in equilibrium after neutrino decoupling, the decay products could then lead to a measurable $\Delta N_{\rm eff}$ in CMB.
It could also affect BBN if the production of light elements is not concluded yet.
Therefore, observational constraints generally require $\phi$ to be non-relativistic when BBN starts, \ie~$m_\phi~\gtrsim~10~\rm MeV$.

In our model, using Eq.~(\ref{eq:width_phinunu}), the SM temperature at which the decay occurs is roughly estimated to be
\beqn
T_{\rm dec}&\approx& \frac{3\sqrt{\bar{g}_{\nu}^2+g_{\nu}^2}}{4\sqrt{2}\pi}\left(\frac{10}{g_{\star}(T_{\rm dec})}\right)^{1/4} \left(\frac{m_{\phi}}{\rm MeV}\right)^{1/2}\left(\frac{M_{P}}{\rm MeV}\right)^{1/2}~{\rm MeV}\nn\\
&\approx&  10^{10} \times\sqrt{\bar{g}_{\nu}^2+g_{\nu}^2}\left(\frac{10}{g_{\star}(T_{\rm dec})}\right)^{1/4} \left(\frac{m_{\phi}}{\rm MeV}\right)^{1/2}~{\rm MeV}\,\,\,.
\eeqn
Together with the relic abundance estimate in Eq.~(\ref{eq:yield_approx}), we see for $m_\chi~\sim~\mathcal{O}({\rm GeV})$ and $g_{\chi}$ or $\bar g_\chi~\lesssim~\mathcal{O}(1)$, $T_{\rm dec}~\approx~ (m_\phi/{\rm MeV})^{1/2}~\rm MeV$.
We have estimated that in order to have an out-of-equilibrium decay before BBN i.e. $m_\phi~ >~T_{\rm dec}~>~10 {\rm MeV}$ or an in-equilibrium decay with $\phi$ becoming non-relativistic when BBN starts, $m_\phi$ roughly needs to be larger than $10$ MeV.
This explains the choice of the masses in our benchmark examples.
Among the benchmark results that we have shown, the cases in FIG.~\ref{fg:purefi} and the cases in the lower two rows of FIG.~\ref{fg:reannihilation} correspond to the in-equilibrium decay, whereas the cases in the upper two rows of FIG.~\ref{fg:reannihilation} correspond to the out-of-equilibrium decay.
In the former cases, our choice of mass $m_\phi~=~50~\rm MeV$ allows $\phi$ to be non-relativistic when BBN starts, therefore its energy density is already negligible.
For the latter cases, the fact that $T_{\rm dec}~>~10~\rm MeV$ ensures that no BBN or CMB bounds would be violated.

\subsection{Indirect Detection}
\begin{figure}[t]\centering
\includegraphics[width=0.6\textwidth]{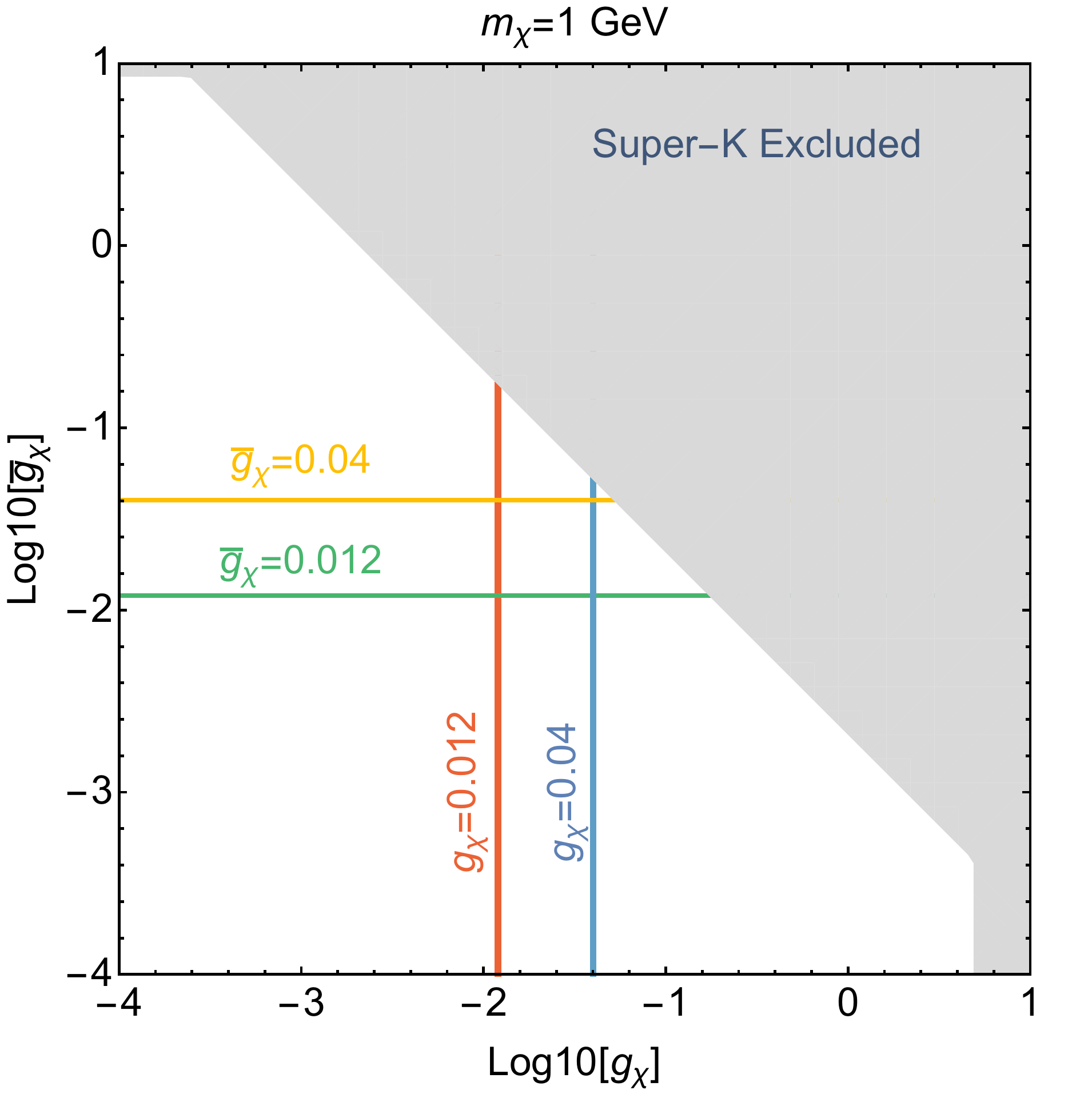}
\caption{We plot the indirect detection constraint from the Super-Kamiokande.
The excluded region is colored in gray.
Different colored lines are related to the four benchmark points in FIG.\,\ref{fg:reannihilation}. 
As the benchmark all have one of the neutrino couplings vanished, they seat at infinitely negative $g_\chi$ or $\bar{g}_\chi$ along the lines in this $\log$-scale plot.   
}\label{fg:indrect}
\end{figure}

The decay of $\phi$ could also be exploited in the indirect detection of dark matter.
In the late-time universe, though the interaction rate between the dark matter $\chi$ the SM particle is extremely small, $\chi\chi\to\phi\phi$ followed by rapid decay of $\phi\to\nu\nu$ could potentially produce a signal that can be probed by astrophysical neutrino observation experiments. 
These experiments usually give bounds on the dark matter annihilation cross-section $\langle\sigma v\rangle$. 
In the non-relativistic regime, $\langle\sigma v\rangle$ can be expanded into powers of the relative velocity $v$:
\begin{eqnarray}
\langle\sigma v\rangle~=~a + b \langle v^2\rangle +{\cal O}(v^4)\,\,\,,
\end{eqnarray} 
where the angle brackets represent the average over the local dark-matter distribution at the signal source.
In our model $a$ and $b$ can be expressed as:
\begin{eqnarray}
a &=& \frac{g_\chi^2\bar{g}_\chi^2m_\chi \sqrt{m_\chi^2-m_\phi^2}}{4\pi(2m_\chi^2-m_\phi^2)^2}\,\,\,,\\
b &=&\frac{m_\chi}{24\pi(2m_\chi^2-m_\phi^2)^4\sqrt{m_\chi^2-m_\phi^2}} \Big\{ g_\chi^4(18 m_\chi^6+20m_\phi^4m_\chi^2-34m_\phi^2 m_\chi^4 -4m_\phi^6)  \nonumber \\
&~& +\big[2\bar{g}_\chi^4(m_\chi^2-m_\phi^2)^3+3g_\chi^2 \bar{g}_\chi^2(m_\phi^6 - 8 m_\phi^4m_\chi^2 +20m_\phi^2 m_\chi^4-12 m_\chi^6)\big] \Big\}\,\,\,.
\end{eqnarray}
According to the result in Ref.~\cite{Arguelles:2019ouk}, we find that the most stringent bound on the annihilation cross-section $\langle\sigma_{\chi\chi\to \nu\nu}v\rangle$ for $m_{\chi}~\sim~1~\rm GeV$  comes from Super-Kamiokande (SK) experiment \cite{Fukuda:2002uc,Richard:2015aua}, which translates into  $a~\lesssim~10^{-24}~\rm\ cm^3\cdot s^{-1}$ and $b~\lesssim~10^{-16}~\rm\ cm^3\cdot s^{-1}$. 
We plot this constraint against $\bar{g}_\chi$ and $g_\chi$ in FIG.~\ref{fg:indrect} where the gray region is excluded by the SK experiment. 
Our benchmark points in FIG.\,\ref{fg:reannihilation} all seat infinitely far away along the four colored lines which cannot be shown on the $\log$-scale plot as they all have one of the neutrino couplings $g_\chi$ or $\bar{g}_\chi$ being zero. 
Nevertheless, our benchmarks all survive from the indirect detection constraint.

\FloatBarrier

\subsection{Possible Solution to the Small Scale Structure Problem}
\begin{figure}[t]\centering
\includegraphics[width=0.6\textwidth]{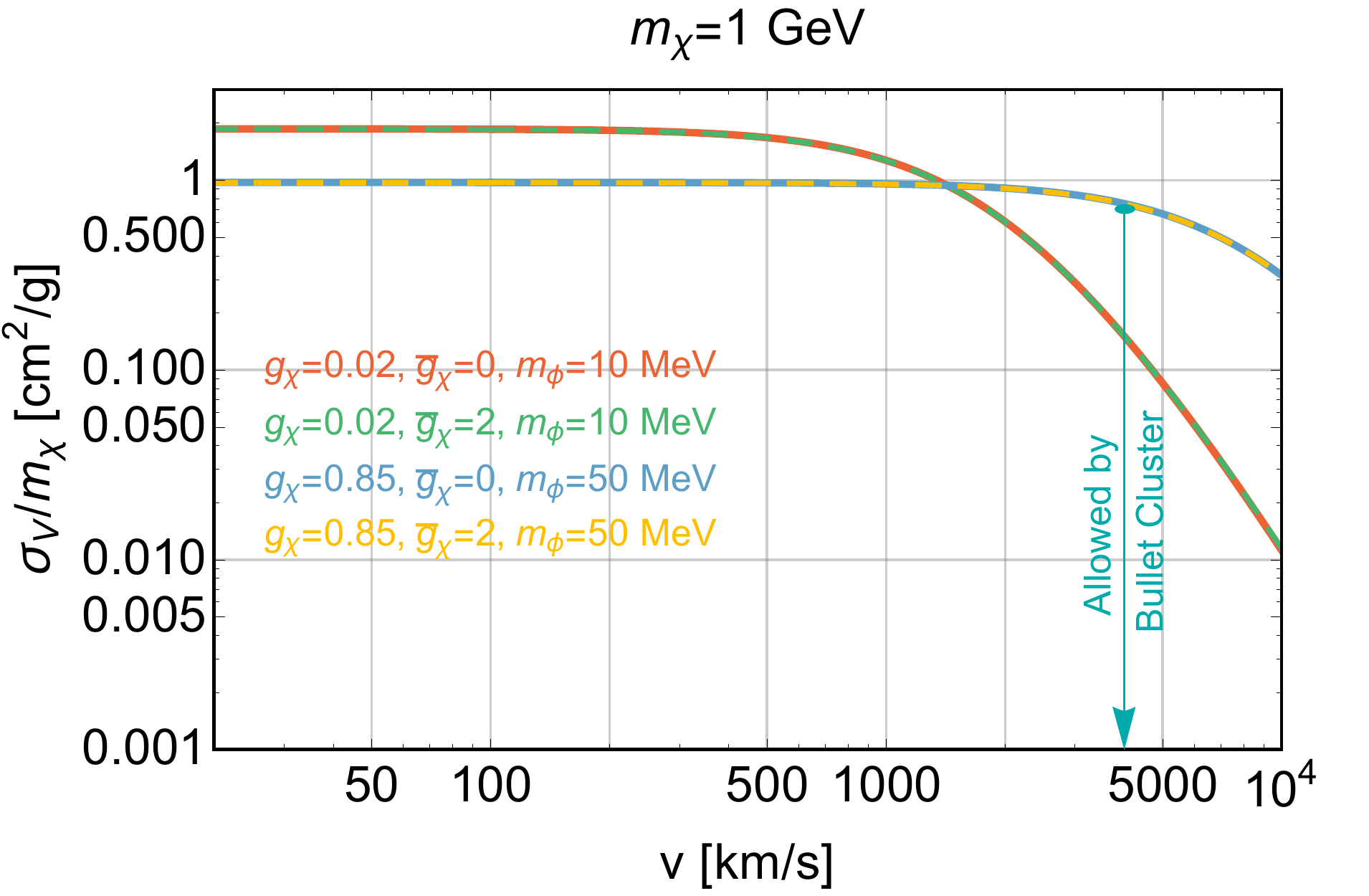}
\caption{Viscosity cross-section $\sigma_V$ per dark mass $m_\chi$ versus the velocity. 
Four curves are clustered in two groups, $m_\phi=10~\rm MeV$ and $m_\phi=50~\rm MeV$.
The values of $g_\chi$ are chosen such that at low velocity $\sigma_V/m_\chi$ is ${\cal O}(1)~{\rm cm^2/g}$.
The values of $\bar{g}_\chi$ are chosen to demonstrate its negligible effect on the $\sigma_V/m_\chi$.
The cyan arrow represents the constraint from the bullet cluster: $\sigma_V/m_\chi~\lesssim~0.7~{\rm cm^2/g}$ for $v~\sim~4000~{\rm km/s}$ \cite{Tulin:2017ara}. 
}\label{fg:selfxsec}
\end{figure}
Finally, we investigate the potential for our model to solve problems on small scale structures.
It is well known that though the standard $\Lambda$CDM cosmology is very successful in predicting the large scale structure of the universe,  
it fails to predict correct structures that are compatible with the observational data at the small scales.
The most prominent small-scale problems are the follows: 
1) \textit{the core-cusp problems} --- the dark-matter density profile in the center of a galaxy predicted from simulations with collisionless cold dark matter (CDM) typically scales as $r^{-1}$ (``cuspy''), while the observations show that both cored (dark matter density $\sim~r^0$) and cuspy centers exist \cite{Kahlhoefer:2019oyt};
2) \textit{the missing satellites problem} --- the observed number of satellite galaxies in the local group is one to two orders of magnitude fewer than the prediction from CDM simulations;
\textit{the too-big-to-fail problem} --- when matching the most luminous observed satellites of the Milky Way with the most massive subhalos from CDM simulations, the subhalos appear to be too massive to host those observed galaxies\footnote{This is the too-big-to-fail problem in the Local Group. 
The too-big-to-fail problem in the field refers to the fact that the rotation curves for a large fraction of the dwarf galaxies indicates that they reside in halos with masses that are smaller than what is predicted using abundance matching between the observation and the CDM simulations \cite{Papastergis:2014aba} --- implying that some of the more massive halos fail in forming their galactic counterpart.}.
These problems motivate the consideration on the self-interaction of dark-matter particles which could potentially solve the problem by lowering the density in the central region of a galaxy as compared to the CDM simulations.
An excellent review on this topic is provided by {the authors of} Ref.\ \cite{Tulin:2017ara}, where a rough bound on the dark-matter self-scattering cross-section over its mass $\sigma/m_{\chi}$ can be derived by fitting the astrophysical data. 
As pointed out in Ref.\ \cite{Tulin:2017ara}, $\sigma/m_{\chi}~\sim~\mathcal{O}(1)\ {\rm cm^2/g}$ can solve the small-scale problems, and a transition in the scaling of the velocity dependence of the self-interacting cross-section around ${\cal O}(10^3)~\rm km/s$ is crucial
for reconciling the collisionless nature of the dark matter on large scales with the collisional behavior on small scales. 
On the other hand, Ref.\ \cite{Tulin:2013teo} points out that the viscosity cross section
\beqn
\sigma_V~\equiv~2\pi \int_0^\pi \frac{d\sigma}{d\Omega}(1-\cos\theta^2) \sin\theta {\rm d}\theta \,\,\,,
\eeqn
is a better proxy than the center-of-mass cross-section $\sigma$ in characterizing the self-interacting strength of the dark matter when dealing with small scale structure problems as it emphasizes the momentum transfer in the perpendicular direction and better {describes} the speed of energy equalization.
Therefore, we calculate $\sigma_V/m_{\chi}$ in our model 
and find that, at zero velocity, to leading order in the expansion with respect to $m_\phi/m_\chi$
\beq
\sigma_V~\sim~\frac{g_\chi^4 m^2_\chi}{6\pi m^4_\phi}\,\,\,.\label{eq:sigmaV_exp}
\eeq
Notice that it depends on $g_{\chi}$ only as the terms containing $\bar g_\chi$ are suppressed by at least $(m_\phi/m_\chi)^2$.
This makes it more difficult if not entirely impossible for scenarios with $\bar g_\chi$ alone to solve the problems on small scale structures.
Our result of $\sigma_V/m_\chi$ is shown in FIG.~\ref{fg:selfxsec}, where we take four benchmark points with fixed $m_\chi~=~1~\rm GeV$. 
Two values of $m_\phi$, $10$ MeV and $50$ MeV, are selected to illustrate the difference in the velocity dependence. 
For each $m_\phi$, we further select a value for $g_\chi$ to ensure that $\sigma_V/m_\chi~\sim~{\cal O}(1)\ {\rm cm^2/g}$ at low velocity in order to solve the small scale structure problem, and two extreme values for $\bar{g}_\chi$ to demonstrate its weak effect on $\sigma_V$. 

Obviously, for the benchmark masses we select in FIG.\,\ref{fg:reannihilation}, ${\cal O}(1)$ value is needed for $g_{\chi}$ in order to solve the small scale problems while roughly satisfying the constraints on large scales (\eg~the Bullet Cluster constraint).
However, a $g_{\chi}$ of this size would lead to a long period of dark thermal equilibrium or QSE, making the yield of dark matter too low at the dark freeze-out.
In addition, since $g_\nu$ and/or $\bar{g}_\nu$ are bounded from above by the requirement on freeze-in production, increasing $g_\nu$ and/or $\bar{g}_\nu$ could not solve the problem with our choice of mass parameters. 
On the other hand, decreasing $m_\phi$ to $\sim~10~\rm MeV$ could allow a smaller $g_\chi$ and is thus able to solve the small scale problems and obtain the correct relic abundance simultaneously.
However, for suitable values of $g_\nu$ and $\bar{g}_\nu$ (roughly the same order as the values shown in FIG.\,\ref{fg:purefi} and FIG.~\ref{fg:reannihilation} since it is not sensitive to $m_\phi$) that yields the correct relic abundance, the decay of $\phi$ would violate the BBN constraint by presenting a sizable $\Delta N_{\rm eff}$.
Thus, we find the only way to make our model a potentially viable SIDM candidate while satisfying the constraints on the relic abundance and BBN is to increase $m_{\chi}$ while adjusting $g_\chi$ accordingly to maintain an appropriate $\sigma_V/m_\chi$, since a larger $m_\chi$ would enable $\chi$ to freeze out earlier before getting severely Boltzmann suppressed.

In general, we see that the conditions for solving the small scale structure problems (requiring $g_{\chi}~\sim~\mathcal{O}(1)$ and preferring lighter $m_{\phi}$) tend to act against the requirement to produce the correct dark-matter relic abundance in the freeze-in regime (favoring smaller $g_{\chi}$ to prevent a late dark freeze-out) and the bounds on the decay of $\phi$ (favoring $m_\phi~\gtrsim~10~\rm MeV$).

\FloatBarrier

\section{Conclusion}\label{sec:con}

In this paper, we study a simplified model describing the neutrino-portal dark matter $\chi$.
In this model, the dark matter $\chi$ does not couple directly to the SM particles.
Instead it interacts with the SM neutrinos via an $s$-channel exchange of the light scalar mediator $\phi$. 
Such neutrino-portal interaction could naturally give rise to a tiny coupling between the dark and the SM sector by relating the smallness of the coupling to the smallness of the neutrino mass, and therefore is appropriate for the freeze-in production of dark matter which requires an extremely small interaction rate between the two sectors.
We then study the possible UV completions of our model and the production of dark matter in the early universe focusing on interesting dynamics that could arise in the dark sector.
We also investigate several potential phenomenological constraints associated with our model. 
Our main results are summarized in the following:
\begin{itemize}
\item We point out two UV origins of the simplified model, namely, the type-I seesaw and the Majoron model, that can naturally induce a small pure scalar and pure pseudo-scalar coupling between $\phi$ and neutrinos, respectively. 
\item We study the freeze-in production of $\chi$ in the pure freeze-in regime as well as in the reannihilation regime.
For the former, we analytically estimate the relic abundance and present two benchmark results in FIG.~\ref{fg:purefi} by solving the Boltzmann equation for the number density $n_{\chi}$ numerically.
For the latter, we study the dark-sector dynamics in detail and further identify two different scenarios --- the one with the dark thermal equilibrium only and the one with the additional QSE phase.
Typically, the one with the QSE phase requires a larger rate of injection from the SM sector (larger $g_\nu$ and/or $\bar g_\nu$). 
We find that since the Boltzmann equation for the $n_\chi$ depends on the dark-sector temperature $T_D$, we need to solve the Boltzmann equation for the dark-sector energy density $\rho_D$ in order to obtain the evolution of $T_D$, and use $T_D$ to solve the Boltzmann equation for $n_\chi$.
We present the other four benchmark results which include both scenarios for both the scalar and the pseudoscalar case.
We also present and analyze the evolution of $T_D$ in these cases.
\item We find that the decay of $\phi$ could potentially be constrained by BBN and CMB.
Nevertheless, we point out that, for our benchmark results, a relatively heavy mediator with $m_\phi~\sim~50~\rm MeV$ could avoid the BBN and CMB constraints while simultaneously give the correct relic abundance.
\item The Super-Kamiokande experiment is possible to constrain our model by searching for dark matter annihilating to mediator $\phi$ followed by $\phi$ decaying to SM neutrinos. 
We show in FIG.~\ref{fg:indrect} that our benchmarks are all allowed by the Super-Kamiokande experiment.
\item Since our model could allow sufficiently large self-interaction between dark-matter particles, we analyze the potential for our model to solve the small scale structure problems.
We find since there is a tension between the required properties for a successful SIDM candidate and the constraints on both BBN and the relic abundance, the choice of mass parameters that we choose in the benchmarks cannot solve the problems on small scale structures.
Nevertheless, it is possible to reconcile the problem by increasing the mass of $\chi$.
\end{itemize}

Finally, we would like to point out several caveats.
First of all, in our examples for UV completion, we have not considered the possibility that the heavy right-handed neutrino $N$ could also be a dark-matter candidate.
Such possibility in general depends on the evolution history of the universe at high scales.
We argue that, in scenarios in which the temperature of the universe has never been higher than $M_N$, or the mass of the inflaton is smaller than $M_N$, one would not expect a non-negligible population of $N$.
Our analysis thus implicitly assumed an appropriate cosmological history before the scales relevant for the production of dark matter.
Scenarios in which both $N$ and $\chi$ could play the role of dark matter are nevertheless worth studying.

Second, although we have confined ourselves within the freeze-in production of dark matter through the neutrino portal, if the couplings $g_\nu$ and/or $\bar g_\nu$ are large enough, the dark-matter particles could thermalize with the SM neutrinos and undergo thermal freeze-out from the SM thermal bath.
However, we do not consider this scenario in our paper as we are motivated by the possibility that the interactions between the dark matter and the SM particles are extremely feeble.

Third, as we have pointed out before, in the reannihilation regime, the interaction rate between the dark-sector particles depends on the dark-sector temperature $T_D$.
However, $T_D$ is only physical when the $\chi$ and $\phi$ particles are in thermal equilibrium.
In our analysis, we obtain $T_D$ from the energy density of the dark sector and use it to evaluate the dark-sector interaction rate throughout.
Therefore, our approach relies on the assumption that the thermally averaged cross-sections evaluated at $T_D$ is a good approximation for the actual interaction strength near but beyond the dark freeze-out.
Far away from the dark freeze-out, the interaction within the dark sector becomes negligible, and the evolution of the dark matter number density is dominated by the injection from the SM sector and the Hubble expansion.

{Moreover, the $\nu-\nu-\phi$ interaction is subject to constraints from supernovae as the scalar $\phi$ might be produced copiusly through inverse decay of SM neutrinos in the core of a supernova \cite{Choi:1987sd,Farzan:2002wx,Acharya:2009zt,Heurtier:2016otg,deGouvea:2019qaz}.
However, this constraint is sensitive to both the mass of the scalar and the flavor dependence of the coupling.
Therefore, in principle, such constraint could be evaded by either adjusting $m_\phi$ or the flavor structure of $g_\nu$ and/or $\bar g_\nu$.
A detailed study on this would help identify the viable parameter space for our model, but is beyond the scope of our paper.}

Furthermore, when calculating the thermally averaged cross-sections and the energy transfer rates, we have assumed that the dark-sector particles $\chi$ and $\phi$ all follow a thermal distribution.
While this is true when the dark sector is in thermal equilibrium, it might not be true if the dark sector is not, as the self-scattering rate might be too small to reach kinetic equilibrium.
Therefore, it is possible for the phase-space distributions of $\chi$ and $\phi$ to be highly nonthermal as the SM thermal bath keeps injecting particles into the dark sector while the existing dark-sector particles are constantly redshifting.
This would in principle affect the calculation of the collision terms.
The nonthermal features in the phase-space distribution could also leave observable imprints on the matter power spectrum \cite{Konig:2016dzg,Murgia:2017lwo,Dienes:2020bmn} if some of the $\chi$ particles could have non-negligible momenta during the structure formation.
A thorough study on these effects would require solving the Boltzmann equations at the level of the phase-space distribution for both $\chi$ and $\phi$ which we leave for future work.

\acknowledgments
We would like to thank J.\ Heeck for useful discussions.
YD acknowledges support from U.S. Department of Energy under contract No.\ DE-SC0011095. 
FH is supported by the National Natural Science Foundation of China (NSFC) under grant
No.\ 11947302, No.\ 11690022, No.\ 11851302, No.\ 11675243 and No.\ 11761141011 and also supported by the Strategic Priority Research Program of the Chinese Academy of Sciences under grant No.\ XDB21010200 and No.\ XDB23000000. 
J.\ H.\ Y.\ is supported by the National Science Foundation of China under Grants No.\ 11875003 and No.\ 11947302. 
H.\ L.\ L. is supported by the National Science Foundation of China under Grants No.\ 11875003 and 2019 International Postdoctoral Exchange Fellowship Program.

\appendix

\section{Cross Sections}
In this section, we present relevant cross-sections and decay widths that we have used in this paper based on the general effective Lagrangian in Eq.~(\ref{eq:lag_int}).
For all the practical purposes, we shall treat the SM neutrinos as massless.\\
\\
\textbullet~$\chi\chi\rightarrow \nu\nu$
\beqn
\sigma_{\chi\chi\rightarrow \nu\nu}~=~\frac{3(\bar{g}_{\nu}^2+g_{\nu}^2)s^{1/2}\left[g_{\chi}^2(s-4m_{\chi}^2)+\bar{g}_{\chi}^2s\right]}{32\pi \sqrt{s-4m_{\chi}^2}\left[(s-m_{\phi}^2)^2+m_{\phi}^2\Gamma_{\phi}^2\right]}\,.\label{eq:xsection_xxnunu}
\eeqn
\\
\textbullet~$\phi\phi\rightarrow \nu\nu$
\beqn
\sigma_{\phi\phi\rightarrow \nu\nu}~=~\frac{3(\bar{g}_{\nu}^2+g_{\nu}^2)^2}{8\pi s(s-4m_{\phi}^2)}\Bigg\{\frac{(s^2-4m_{\phi}^2s+2m_{\phi}^4)}{s-2m_{\phi}^2} \ln A-3\sqrt{s(s-4m_{\phi}^2)}  \Bigg\}\,,\label{eq:xsection_phiphinunu}
\eeqn
where we have defined 
\beq
A~\equiv~\frac{s-2m_{\phi}^2+\sqrt{s(s-4m_{\phi}^2)}}{s-2m_{\phi}^2-\sqrt{s(s-4m_{\phi}^2)}}
\eeq
in order to simplify the expression.\\
\\
\textbullet~$\chi\chi\rightarrow \phi\phi$
\beqn
\sigma_{\chi\chi\rightarrow \phi\phi}&=&\frac{1}{32\pi s(s-4m_{\chi}^2)}\nn\\
&~&\times\Bigg\{\frac{\ln B}{s-2m_{\phi}^2} \Big[
(g_{\chi}^2+\bar{g}_{\chi}^2)^2s^2 -4(g_{\chi}^2+\bar{g}_{\chi}^2)\left( (g_{\chi}^2+\bar{g}_{\chi}^2)m_{\phi}^2-4g_{\chi}^2m_{\chi}^2 \right)s\nn\\
&~&+6(g_{\chi}^2+\bar{g}_{\chi}^2)^2m_{\phi}^4 - 16g_{\chi}^2(g_{\chi}^2+\bar{g}_{\chi}^2)m_{\phi}^2m_{\chi}^2-32g_{\chi}^4m_{\chi}^2
\Big]
\nn\\
&~&
-\frac{\sqrt{(s-4m_{\chi}^2)(s-4m_{\phi}^2)}}{m_{\chi}^2s-4m_{\phi}^2m_{\chi}^2+m_{\phi}^4}\Big[g_{\chi}^4\left(2m_{\chi}^2(s+8m_{\chi}^2) -16m_{\phi}^2m_{\chi}^2 +3m_{\phi}^4\right) \nn\\
&~&+\bar{g}_{\chi}^4(2m_{\chi}^2s -8m_{\phi}^2m_{\chi}^2 +3m_{\phi}^4)+2g_{\chi}^2\bar{g}_{\chi}^2(2m_{\chi}^2s -12m_{\phi}^2m_{\chi}^2 +3m_{\phi}^4)
\Big]
\Bigg\}\,,\label{eq:xsection_chichiphiphi}\nn\\
\eeqn
where 
\beq
B~\equiv~ \frac{s-2m_{\phi}^2+\sqrt{(s-4m_{\chi}^2)(s-4m_{\phi}^2)}}{s-2m_{\phi}^2-\sqrt{(s-4m_{\chi})(s-4m_{\phi}^2)}} \,.
\eeq
\\
\textbullet~$\nu\nu\rightarrow \phi$
\begin{equation}
\sigma_{\nu\nu\rightarrow\phi}~=~\frac{3\pi\left(g_{\nu}^2+\bar{g}_{\nu}^2\right)m_{\phi}^2}{4s^{3/2}}\delta(\sqrt{s}-m_{\phi})
\,.\label{eq:xsection_nunuphi}
\end{equation}
\\
\textbullet~$\phi\rightarrow\nu\nu$
\beq
\Gamma_{\phi\rightarrow\nu\nu}~=~\frac{3(g_{\nu}^2+\bar{g}_{\nu}^2)m_{\phi}}{32\pi}\,.\label{eq:width_phinunu}
\eeq

\section{Energy Transfer}\label{sec:energy_transfer}
In this appendix, we present the explicit forms for the energy transfer terms in Eq.\,(\ref{eq:rhodBoltzmann}).
Let us begin with the Boltzmann equation for the phase-space distribution of a particle species $\psi$:
\beq
\frac{\partial f_{\psi}(p_{\psi},t)}{\partial t}~=~H(t)p_{\psi}\frac{\partial f_{\psi}(p_{\psi},t)}{\partial p_{\psi}} + C[f]\,,\label{eq:boltzmann_f}
\eeq
in which $C[f]$ is the collision operator which involves all possible processes for $\psi$, \eg~decay, inverse decay, elastic scattering, annihilation, etc, and thus depends on the phase-space distribution of all relevant species.
For a specific process $\psi+a+b+\dots\leftrightarrow i+j+\dots$, the collision term takes the general form
\beqn
C[f]&=&-\frac{1}{2E_{\psi}}\int d\pi_a d\pi_b\dots d\pi_i d\pi_j\dots (2\pi)^4 \delta^{(4)}(p_{\psi}+p_a+p_b+\dots-p_i-p_j-\dots)\nn\\
&~&\times\bigg[\abs{\mathcal{M}_{\psi+a+b+\dots\leftrightarrow i+j+\dots}}^2f_{\psi}f_af_b\dots(1\pm f_i)(1\pm f_j)\dots\nn\\
&~&~~~~ - \abs{\mathcal{M}_{i+j+\dots\leftrightarrow \psi+a+b+\dots}}^2f_i f_j\dots (1\pm f_{\psi})(1\pm f_a)(1\pm f_b)\dots \bigg]\,,\label{eq:collision}
\eeqn
where $d\pi_i=d^3p_{i}/\left((2\pi)^3 2E_{i}\right)$, and ``$\pm$'' represents the Bose-enhancement/Pauli-blocking effects --- one needs to choose ``$+$'' for bosons whereas ``$-$'' for fermions.
In principle, the phase-space distributions $f$ could take any form.
For species in thermal equilibrium, $f$ is either Bose-Einstein or Fermi-Dirac.
However, for most of the practical purposes, the Maxwell-Boltzmann distribution can be used as a good approximation,
and the condition $f\ll 1$ is often satisfied. 
Therefore, one can use the approximations $f\approx e^{-E/T}$ and $1\pm f\approx 1$ which greatly simplify the calculation.

To get the energy-transfer terms in the Boltzmann equation for the energy density, we just need to multiply Eq.\,(\ref{eq:boltzmann_f}) with $E_{\psi}$ and integrate over the phase-space element $d^3p_{\psi}/(2\pi)^3$.
The result is
\beq
\frac{\partial \rho_{\psi}}{\partial t}~=~-3H(P_{\psi}+\rho_{\psi})+\int \frac{d^3p_{\psi}}{(2\pi)^3} E_{\psi} C[f]\,.
\eeq
For the process $\psi+a+b+\dots\leftrightarrow i+j+\dots$, one can separate the contribution from the forward and backward directions, \ie~the contribution from the two terms in the square brackets in Eq.\,(\ref{eq:collision}).
Thus, the last term can be written formally as
\beq
\int \frac{d^3p_{\psi}}{(2\pi)^3} E_{\psi} C[f]=-n_{\psi}n_a n_b\dots\mathcal{P}_{\psi+a+b+\dots\rightarrow i+j+\dots}+n_i n_j\dots\mathcal{P}_{i+j+\dots\rightarrow \psi+a+b+\dots}\,.\label{eq:energy_transfer_general}
\eeq
In the following subsections, we shall calculate the energy transfer terms for several particular processes.
\\
\\
\textbullet~~Decay: $\phi\rightarrow \nu\nu$\\
\\
The contribution from the decay process $\phi\rightarrow \nu\nu$ is
\beqn
n_{\phi}\mathcal{P}_{\phi\rightarrow\nu\nu}&=&\int d\pi_{\phi} f_{\phi} E_{\phi} \int d\pi_{\nu_1} \int d\pi_{\nu_2}   \abs{\mathcal{M}_{\phi\rightarrow \nu\nu}}^2 
(2\pi)^4\delta^{(4)}(p_{\nu_1}+p_{\nu_2}-p_{\phi})\nn\\
&=&\int \frac{d^3p_{\phi}}{(2\pi)^3}f_{\phi}m_{\phi} \frac{1}{2m_{\phi}}\int d\pi_{\nu_1} \int d\pi_{\nu_2}   \abs{\mathcal{M}_{\phi\rightarrow \nu\nu}}^2 
(2\pi)^4\delta^{(4)}(p_{\nu_1}+p_{\nu_2}-p_{\phi})\nn\\
&=&n_{\phi}m_{\phi} \Gamma_{\phi\rightarrow\nu\nu} \label{eq:decay_E}
\eeqn
where $\Gamma_{\phi\rightarrow\nu\nu}$ is the decay width in the center of mass frame of $\phi$.
\\
\\
\textbullet~~Inverse Decay: $\nu\nu\rightarrow \phi$\\
\\
The energy-transfer term of this process is
\beqn
{n_{\nu}^{\rm eq}(T)}^2\mathcal{P}_{\nu\nu\rightarrow\phi}&=&\int \frac{d^3p_{\nu_1}}{(2\pi)^3}\frac{d^3p_{\nu_2}}{(2\pi)^3}
\sigma_{\nu\nu\rightarrow\phi}v_{\text{M{\o}l}}(E_{\nu_1}+E_{\nu_2}) e^{-E_{\nu_1}/T}e^{-E_{\nu_2}/T}\,.
\eeqn
Here we use the superscript ``eq'' to emphasize that neutrinos are in thermal equilibrium with the rest of the SM thermal bath.
Following the convention in in Gondolo's paper:
\begin{eqnarray}
E_+=E_{\nu_1}+E_{\nu_2}\,,\\
E_-=E_{\nu_1}-E_{\nu_2}\,,\\
E_-^{\min}=-\sqrt{1-4m_{\nu}^2/s}\sqrt{E_+^2-s}\,,\\
E_-^{\max}=\sqrt{1-4m_{\nu}^2/s}\sqrt{E_+^2-s}\,,\\
v_{\text{M{\o}l}}E_{\nu_1}E_{\nu_2}=\frac{1}{2}\sqrt{s(s-4m_{\nu}^2)}\,,
\end{eqnarray}
the above integration becomes
\beqn
&~&\frac{1}{32\pi^4}\int_{4m_{\nu}^2}^{\infty}ds\int_{\sqrt{s}}^{\infty}dE_+\int_{E_{-}^{\min}}^{E_{-}^{\max}}d E_{-}
\sigma_{\nu\nu\rightarrow\phi}v_{\text{M{\o}l}} E_{\nu_1}E_{\nu_2} E_+ e^{-E_{+}/T}\nn\\
&=&\frac{1}{32\pi^4}\int_{4m_{\nu}^2}^{\infty}ds 
\sigma_{\nu\nu\rightarrow\phi}(s-4m_{\nu}^2) \int_{\sqrt{s}}^{\infty}dE_+
\sqrt{E_{+}^2-s}~E_+e^{-E_{+}/T}\,.
\eeqn
Let us first integrate over $dE_+$
\begin{eqnarray}
\int_{\sqrt{s}}^{\infty}dE_+ E_+ \sqrt{E_{+}^2-s}~e^{-E_{+}/T}&=&
sTK_2(\sqrt{s}/T)\,.
\end{eqnarray}
Insert the result above, we obtain the final expression
\beq
n_{\nu}^{\rm eq}(T)^2\mathcal{P}_{\nu\nu\rightarrow\phi}~=~\frac{1}{32\pi^4}\int_{4m_{\nu}^2}^{\infty}ds~ 
\sigma_{\nu\nu\rightarrow\phi}s(s-4m_{\nu}^2) TK_2(\sqrt{s}/T)\,.
\eeq
Notice that this would be exactly the same with $n_{\phi}(T_{\phi})\mathcal{P}_{\phi\rightarrow\nu\nu}$ in Eq.\,(\ref{eq:decay_E}) if $\phi$ is in thermal equilibrium with the SM thermal bath.
\\
\\
\textbullet~~ 2-to-2 process\\
\\
Let us consider the 2-to-2 process $3+4\leftrightarrow 1+ 2$.
The energy-transfer collision term associated with this process for particle $1$  is
\beqn
n_3n_4\mathcal{P}_{34\to 12}-n_1n_2\mathcal{P}_{12\to 34}&=&\int \prod_{i=1}^{4} d\pi_i~(2\pi)^4\delta^{(4)}(p_1+p_2-p_3+p_4)\nn\\
&~&~~\times E_1\left[f_3(p_3)f_4(p_4)\overline{\abs{\mathcal{M}_{3 4 \rightarrow 1 2}}^2}-f_1(p_1)f_2(p_2)\overline{\abs{\mathcal{M}_{1 2 \rightarrow 3 4}}^2}\right]\,,\nn\\\label{eq:2to2_general}
\eeqn
where the ``bar'' over the amplitude $\abs{\mathcal{M}}^2$ indicates necessary average over different spin states.
For our purpose, we assume that particles $3$ and $4$ are the states in the SM thermal bath, whereas particles $1$ and $2$ are dark-sector states.
Since for freeze-in process, the energy transferred from the dark sector to the visible sector is negligible, we just need to calculate the energy transferred to the dark sector through the above process.
Therefore, we just need to calculate the forward process --- the part associated with the first term in the brackets:
\beqn
n_3^{\rm eq}(T)n_4^{\rm eq}(T)\mathcal{P}_{34\to 12}&=&\int \prod_{i=1}^{4} d\pi_i~(2\pi)^4\delta^{(4)}(p_1+p_2-p_3+p_4)\overline{\abs{\mathcal{M}_{3 4 \rightarrow 1 2}}^2} E_1 f_3^{\rm eq}(p_3)f_4^{\rm eq}(p_4)\,\nn\\.
\eeqn
Using detailed balance $f_3^{\rm eq}(p_3)f_4^{\rm eq}(p_4)=f_1^{\rm eq}(p_1)f_2^{\rm eq}(p_2)$ and unitarity, and inserting the flux factor $F\equiv [(p_1\cdot p_2)^2-m_1^2 m_2^2]^{1/2}$ \cite{Gondolo:1990dk}, we can rewrite the above integral as
\beqn
&~&\int \frac{d^3p_1}{(2\pi)^3}\frac{d^3p_2}{(2\pi)^3}
f_1^{\rm eq}(p_1)f_2^{\rm eq}(p_2)
E_1\frac{F}{E_1E_2}\nn\\
&~&~~\times\frac{1}{4F}\int\frac{d^3p_3}{(2\pi)^3 2E_3}\frac{d^3p_4}{(2\pi)^3 2E_4}(2\pi)^4\delta^{(4)}(p_1+p_2-p_3+p_4)\overline{\abs{\mathcal{M}_{1 2 \rightarrow 3 4}}^2}\nn\\
&=&\int \frac{d^3p_1}{(2\pi)^3}\frac{d^3p_2}{(2\pi)^3}
f_1^{\rm eq}(p_1)f_2^{\rm eq}(p_2)
E_1\sigma_{12\to 34}v_{\text{M{\o}l}}\,.\label{eq:2to2_a}
\eeqn
in which we have used the fact that $v_{\text{M{\o}l}}=F/E_1E_2$, and the second line is simply $\sigma_{12\to 34}$.
Comparing with the second term in the square brackets of Eq.\,(\ref{eq:2to2_general}), it is easy to notice that the calculation above is exactly equivalent to calculating the energy transfer rate of the inverse process $1+2\to 3+4$ if particle $1$ and $2$ are in thermal equilibrium with the visible sector.
This observation proves that 
\beq
n_3^{\rm eq}(T)n_4^{\rm eq}(T)\mathcal{P}_{34\to 12}=n_1^{\rm eq}(T)n_2^{\rm eq}(T)\mathcal{P}_{12\to 34}\,,
\eeq
which allows us to trade the energy-transfer term of one process for that of its inverse.
Following the procedure in Ref.~\cite{Krnjaic:2017tio}, the Eq.\,(\ref{eq:2to2_a}) can be further simplified, and we arrive at this final expression:
\beq
n_3^{\rm eq}(T)n_4^{\rm eq}(T)\mathcal{P}_{34\to 12}~=~\frac{T}{64\pi^4}\int_{s_0}^{\infty} ds~\sigma_{12\to 34}  s(s-s_0) K_2\left( \sqrt{s}/T \right)\,,
\eeq
where $s_0$ is the minimum of $s$.

\bibliographystyle{JHEP}
\bibliography{ref}

\end{document}